\documentclass[twocolumn]{aastex62}

\usepackage{amsmath, mathtools, amsthm, amssymb}
\usepackage{enumitem}
\usepackage{footmisc}
\usepackage{lineno}
\hypersetup{colorlinks=true, allcolors=cyan}

\received{XX}
\revised{YY}
\accepted{ZZ}
\submitjournal{ApJ}

\shorttitle{Star Formation Quenching in Groups}
\shortauthors{Rhee et al.}

\def\NH{\textsc{NewHorizon}}
\def\VR{\textsc{VELOCIraptor}}
\def\HAGN{\textsc{Horizon-AGN}}
\def\ramses{\textsc{Ramses}}

\newcommand{\HH}{\textsc{H}\textsubscript{2}}

\begin{document}
\newcommand\edel{\bgroup\markoverwith
{\textcolor{red}{\rule[0.5ex]{2pt}{0.8pt}}}\ULon}
\newcommand{\ecom}[1]{{\color{orange}{{``\it #1"}}}}
\newcommand{\eadd}[1]{{\color{red}{{#1}}}}
\newcommand{\sky}[1]{{\color{blue}{{[SKY: #1]}}}}

\let\clearpage\relax
\title{On the Origin of Star Formation Quenching of Galaxies in Group Environments using the NewHorizon simulation}

\email{jinsu.rhee@gmail.com}
\email{yi@yonsei.ac.kr}

\author{Jinsu Rhee}
\affil{Department of Astronomy and Yonsei University Observatory, Yonsei University, Seoul 03722, Republic of Korea}
\affil{Korea Astronomy and Space Science Institute, 776, Daedeokdae-ro, Yuseong-gu, Daejeon 34055, Republic of Korea}

\author{Sukyoung K. Yi}
\affil{Department of Astronomy and Yonsei University Observatory, Yonsei University, Seoul 03722, Republic of Korea}

\author{Jongwan Ko}
\affil{Korea Astronomy and Space Science Institute, 776, Daedeokdae-ro, Yuseong-gu, Daejeon 34055, Republic of Korea}
\affil{University of Science and Technology, Gajeong-ro, Daejeon 34113, Republic of Korea}


\author{Emanuele Contini}
\affil{Department of Astronomy and Yonsei University Observatory, Yonsei University, Seoul 03722, Republic of Korea}

\author{J. K. Jang}
\affil{Department of Astronomy and Yonsei University Observatory, Yonsei University, Seoul 03722, Republic of Korea}

\author{Seyoung Jeon}
\affil{Department of Astronomy and Yonsei University Observatory, Yonsei University, Seoul 03722, Republic of Korea}

\author{San Han}
\affil{Department of Astronomy and Yonsei University Observatory, Yonsei University, Seoul 03722, Republic of Korea}

\author{Christophe Pichon}
\affil{Institut d'Astrophysique de Paris, CNRS, Sorbonne Universit\'{e}, UMR 7095, 98 bis bd Arago, 75014 Paris, France}
\affil{IPhT, DRF-INP, UMR 3680, CEA, L'orme des Merisiers, B\^{a}t 774, 91191 Gif-sur-Yvette, France}
\affil{Korea Institute for Advanced Study, 85 Hoegi-ro, Dongdaemun-gu, Seoul 02455, Republic of Korea}

\author{Yohan Dubois}
\affil{Institut d'Astrophysique de Paris, CNRS, Sorbonne Universit\'{e}, UMR 7095, 98 bis bd Arago, 75014 Paris, France}

\author{Katarina Kraljic}
\affil{Observatoire Astronomique de Strasbourg, Universit\'e de Strasbourg, CNRS, UMR 7550, F-67000 Strasbourg, France}

\author{S\'{e}bastien Peirani}
\affil{Institut d'Astrophysique de Paris, UMR 7095, 98 bis Boulevard Arago, F-75014 Paris, France}
\affil{Universit\'e C\^ote d'Azur, Observatoire de la C\^ote d'Azur, CNRS, Laboratoire Lagrange, Bd de l'Observatoire, \\
CS 34229, 06304 Nice Cedex 4}
\affil{Department of Physics, School of Science, The University of Tokyo, 7-3-1 Hongo, Bunkyo-ku, Tokyo 113-0033, Japan}


\begin{abstract}
We study star formation (SF) quenching of satellite galaxies with $M_{*} > 10^7\,M_{\odot}$ within two low-mass groups ($M_{\rm vir}=10^{12.9}$ and $10^{12.7} \,M_{\odot}$) using the NewHorizon simulation.
We confirm that satellite galaxies ($M_{*}\lesssim10^{10}\,M_{\odot}$) are more prone to quenching than their field counterparts.
This quenched fraction decreases with increasing stellar mass, consistent with recent studies.
Similar to the findings in cluster environments, we note a correlation between the orbital motions of galaxies within these groups and the phenomenon of SF quenching.
Specifically, SF is suppressed at the group center, and for galaxies with $M_{*} > 10^{9.1}\,M_{\odot}$, there is often a notable rejuvenation phase following a temporary quenching period.
The SF quenching at the group center is primarily driven by changes in star formation efficiency and the amount of gas available, both of which are influenced by hydrodynamic interactions between the interstellar medium and surrounding hot gas within the group.
Conversely, satellite galaxies with $M_{*} < 10^{8.2}\,M_{\odot}$ experience significant gas removal within the group, leading to SF quenching.
Our analysis highlights the complexity of SF quenching in satellite galaxies in group environments, which involves an intricate competition between the efficiency of star formation (which depends on the dynamical state of the gas) on the one hand, and the availability of cold dense gas on the other hand.
This challenges the typical understanding of environmental effects based on gas stripping through ram pressure, suggesting a need for a new description of galaxy evolution under mild environmental effects.                                                                                                                                                                                                                                                                                                                                                                                                                                                                                                                                                                                                                                                                                                                                                                                                                                                                                                                                                                                                                                                                                                                                                                                                                                                                                                                                                                                                                                                                                                                                                                                                                                                                                                                                                                                                                                                                                                                                                                                                                                                            
\end{abstract}

\section{Introduction}
\label{sec:Intro}

The existence of two distinct populations of galaxies, passive and star-forming, has been reported since the advent of large-scale surveys.
Examples can be observed on the color-magnitude plane of galaxies, with a sequence of red passive galaxies and a cloud of blue star-forming galaxies, dubbed the ``red sequence'' and the ``blue cloud,'' respectively \citep[][]{Strateva01, Bell04, Baldry04}.

The presence of these two populations of galaxies has been observed in various environments \citep[][]{Hogg03, Balogh04, Blanton05} and redshifts \citep[][]{Bell04, Faber07, Mei09}.
Rather than being solely attributed to intrinsic differences, it is known that galaxies in the blue cloud transition to the red sequence through the green valley that lies between these two regions \citep[][]{Faber07, Martin07, Salim07, Schawinski14}.
This naturally leads to the question of the origin of these two populations and the underlying physical mechanisms driving the journey of galaxies on the color-magnitude plane.


Among the various explanations reported so far, the two most prominent mechanisms behind such transitions are associated with the mass and environments of galaxies \citep[e.g.,][]{Balogh04, Kauffmann04, Kimm09, Peng10}, commonly referred to as mass quenching and environmental quenching \citep[][]{Peng10}.
Red and dead phases are preferred by galaxies with higher stellar mass and in denser environments.
Considering that the mass and local density of galaxies evolve over time, this implies that both factors intricately contribute to the formation of passive galaxies \citep[e.g.,][]{Contini19, Rhee20, Jeon22}.

Mass quenching involves internal processes of galaxies, including supernova (SN) feedback at the low-mass end \citep[][]{Larson74, DS86, Benson03, CO06} and active galactic nucleus (AGN) feedback at the high-mass end \citep[][]{Croton06, Silverman08}.
Both feedback mechanisms serve as self-regulating processes in the baryon cycle, where an increase in star formation (SF) triggers a feedback loop involving processes such as heating and dispersion within the interstellar medium (ISM).
For galaxies with $M_{*} > 10^{9} \, M_{\odot}$, the efficiency of mass quenching decreases with decreasing galaxy mass \citep[e.g.,][]{Peng10}.
Additionally, a study by \cite{Geha12} reported no quenched field galaxies with $M_{*} < 10^{9} \, M_{\odot}$, suggesting that the low efficiency of mass quenching extends to even lower mass ranges.
However, recent studies (Sugata et al. in prep) have reported the presence of quenched field galaxies in the low mass regime, indicating that internal quenching processes may affect dwarf galaxies.
Therefore, our current understanding of mass (internal) quenching suggests that it has the weakest impact on galaxies with $M_{*} \sim 10^{9-10} M_{\odot}$.

On the other hand, in extreme environments such as galaxy cluster halos ($M_{\rm vir} >10^{14}\,M_{\odot}$), the existence of passive galaxies is attributed to more dramatic effects.
These are often classified based on the type of interaction forces, either gravitation or hydrodynamic in nature.
Tidal interactions \citep[e.g.,][]{Moore96, Limousin09, Smith16} and ram-pressure stripping \citep[e.g.,][]{GG72, Chung07, LC18} are among the most well-known representatives of these two types of interactions.
A general agreement on environmental quenching for clusters suggests that SF of galaxies with $M_{*} \gtrsim 10^{10}\,M_{\odot}$ are quenched in a delayed-then-rapid manner \citep[][]{Wetzel13, Rhee20, Oman21}: quenching of SF is initially delayed upon infall into clusters, followed by a rapid quenching phase at the pericenter perhaps due to strong ram pressure stripping.
Even though mass quenching and environmental quenching have distinct origins, in most cases, it is not possible to separate the effects of both factors because they simultaneously affect galaxies \citep[e.g.,][]{Darvish16, Contini20, Rhee20}.



However, low-mass galaxies show different trends of mass and environmental quenching from those for galaxies $M_{*} > 10^{10}\,M_{\odot}$.
For example, \cite{Geha12} reported that quenched galaxies with $M_{*} < 10^{9}\,M_{\odot}$ are rare in field regions, suggesting that $M_{*} \sim 10^{9}\,M_{\odot}$ serves as a threshold for mass quenching \citep[see also][]{Haines08, Davies16}.
On the contrary, in the Local Group (LG) as a small group halo in virial mass, many dwarf galaxies are observed to be in a quiescent state \citep[][]{Einasto74,Mateo98, GP09, Tolstoy09, Weisz15, Wetzel15, Putman21}.
The following studies with numerical simulations have proposed rapid quenching scenarios for satellite galaxies in LG-like halos \citep[][]{BM15, Wetzel16, Simpson18, Akins21,Font22}, without the necessity of a delay phase.
This is in line with the prevalence of observed quiescent dwarf galaxies in group halos extending beyond the LG, as indicated by recent studies \citep[][]{Karachentsev13, Davies16}.
However, other studies with SAGA observations \citep[e.g.,][]{Mao21} reported a lower quenched fraction of satellite dwarf galaxies compared to those in LG.
This implies that a consensus on the quenching of low-mass galaxies in low-mass halos has not been reached.

Environmental effects on galaxies in galaxy groups are of utmost importance due to their notably distinct galaxy populations compared to field regions.
In addition, if group halos are eventually accreted into larger clusters, the cumulative effects that occur within these groups are directly connected to pre-processing effects \citep[][]{Mihos04, deLucia12, Han18}.
Therefore, group halos can provide complementary insights into the study of environmental effects.
However, investigating group satellite galaxies through observations has been challenging, primarily because most of them have low luminosity, and hence low surface brightness.
Recently, advancements in observational facilities have enabled high-resolution studies of group satellite galaxies \citep[e.g.,][]{Lee22}.
These observational results have shown remarkably different features of group satellite galaxies (their gas morphology, bulk properties, etc.), revealing new insights into the studies of environmental effects within groups.
In this context, the need for numerical simulations has become increasingly apparent due to the necessity of explaining the origin of such distinct characteristics of group satellite galaxies.

Investigations of such low-mass galaxies in low-mass halos require cosmological simulations with very high mass and spatial resolutions, to resolve low-mass satellite galaxies down to $M_{*} \sim 10^{7} M_{\odot}$ in a live halo with $M_{\rm vir} \sim 10^{12-13}\,M_{\odot}$.
The \NH\ simulation \citep[][]{Dubois21}, a cosmological hydrodynamic zoom-in simulation, satisfies these requirements and hence is one of the best tools for our study.
Using the \NH\ simulation, we address the issue of how differently low-mass satellite galaxies are quenched in low-mass groups ($M_{\rm vir} \lesssim 10^{13}\,M_{\odot}$) to shed light on the detailed mechanism driving their quenching.


\section{Data}
\label{sec:Data}

\subsection{Simulation}
We make use of the two group-size halos from the \NH\ simulation \citep[][]{Dubois21}, a high-resolution follow-up to the \HAGN\ simulation \citep[][]{Dubois14}.
This simulation, conducted using the adaptive mesh refinement code, \ramses\ \citep[][]{Teyssier02}, focuses on a spherical field region with a radius of $10\, {\rm cMpc}$ in the \HAGN\ volume and adopts the WMAP-7 cosmology \citep[][]{Komatsu11}: $\Omega_{m} = 0.272$, $\Omega_{\lambda} = 0.728$, $\sigma_{8}=0.81$, $n_{s} = 0.967$, and $H_{0}=70.4\, {\rm km\,s^{-1}} \, {\rm Mpc^{-1}}$.
The spatial resolution of \NH\ is about 30 times better than that of \HAGN\, reaching the ``best'' resolution of $34\,{\rm pc}$ at $z = 0$ that allows to capture multi-phase ISM.
The mass resolution is $\sim 10^{4} \,M_{\odot}$ for stellar particles and $1.2 \times 10^{6} \,M_{\odot}$ for DM particles, enabling investigations into galaxies down to the dwarf mass regime ($\gtrsim 10^{7}\, M_{\odot}$).
\NH\ has reached the final redshift of $z=0.171$ \footnote{The best spatial resolution at this redshift is $29\,{\rm pc}$ in physical unit.} and consists of 863 consecutive snapshots with a scale factor in a range of $a=0.022$--0.853.
Two adjacent snapshots have approximately $15\,{\rm Myr}$ time interval.
The high mass, spatial, and temporal resolutions of \NH\ are critical to study detailed kinematic properties of disk galaxies and the formation of dwarf galaxies \citep[][]{Jackson21a, Jackson21b, Martin22, Reddish22, Jang23, Yi23}.

The simulation also includes various astrophysics, and here we leave partial details relevant to this analysis.
In \NH , a homogeneous UV background heating occurs after the reionization epoch ($z = 10$), following the Gaussian fitting results from \cite{HM96}.
Cooling processes are modeled for the primordial gas with collisional interactions, recombination, and free-free interaction, allowing it to cool down to $\sim 10^4\,{\rm K}$.
Metal-enriched gas can cool down to $0.1\,{\rm K}$ using the tabulated cooling rates from \cite{DM72} and \cite{SD93}.
Therefore, along with the high-resolution features of \NH , it can depict the multi-phase characteristics of ISM of galaxies.
SF takes place in gas cells with $n_{\rm H} > 10\, {\rm cm}^{-3}$, following the Schmidt law.
However, rather than using a constant value for star formation efficiency (SFE), \NH\ adopts the gravo-turbulent star formation model from \cite{Kimm17} with varying SFE as a function of turbulent Mach number and virial parameter of gas cells (see Section \ref{sec:result-SF} for details), which well reproduces the observed Kennicutt-Schmidt relation \citep[e.g.,][]{Katarina24}.
Feedback from Type II SNe is emitted from newly born star particles.
A star particle in \NH\ can represent a simple stellar population.
\NH\ assume a Chabrier initial mass function \citep[][]{Chabrier05} with cut-off masses of $0.1\, M_{\odot}$ and $150\,M_{\odot}$, resulting in occurring $0.015\, M_{\odot}^{-1}$ SN explosion for each stellar particle and releasing the kinetic energy of $10^{51}\,{\rm erg}$ for each SN explosion.
However, the SN explosion frequency is increased by a factor of 2, and hence $0.03\,M_{\odot}^{-1}$, to compensate for amplified effects from clustered SN explosions that are unresolved in a single stellar particle.
Each SN explosion in \NH\ follows the mechanical SN feedback scheme \citep[][]{KC14, Kimm15}, propagating into the surroundings as a Sedov-Taylor blast wave with energy-conserving (adiabatic) or momentum-conserving (snowplough) phase (see \citealp{Dubois21} for details).
\NH\ does not take feedback from stellar winds and type Ia SNe into consideration.
The simulation also includes super-massive black holes, their accretion, and their feedback, and the detailed specifications (including subgrid physics and certain calibration results) of the simulation can be found in \cite{Dubois21}.

\subsection{Group and satellite galaxies}
The identification of halos and galaxies in the \NH\ simulation is performed using the 6D friends-of-friends galaxy finder code, \VR\ \citep[][]{Elahi19a}, in 834 snapshots with scale factors in a range of $a=0.091$ -- 0.853.
Due to the high particle resolution of \NH, computing the 6D distances between all the particles in massive galaxies (e.g., $>10^{7}$ particles in $M_{*}>10^{11} M_{\odot}$ galaxies), which is required for galaxy and halo detection, takes an impractically long time. 
We use the performance-optimized version of the \VR\ code \citep{Rhee22}\footnote{\href{https://github.com/JinsuRhee/VELOCIraptor-STF}{https://github.com/JinsuRhee/VELOCIraptor-STF}} \footnote{\href{https://github.com/JinsuRhee/NBodylib}{https://github.com/JinsuRhee/NBodylib}}, which significantly improves the efficiency of the full 6D distance computations.
All galaxies and halos are successfully detected in a practically feasible execution time (see \citealp{Rhee22} for details).
Merger trees for all galaxies and halos are constructed using \textsc{TreeFrog} \citep[][]{Elahi19b}.

\begin{figure*}
\centering
\includegraphics[width=0.99\textwidth]{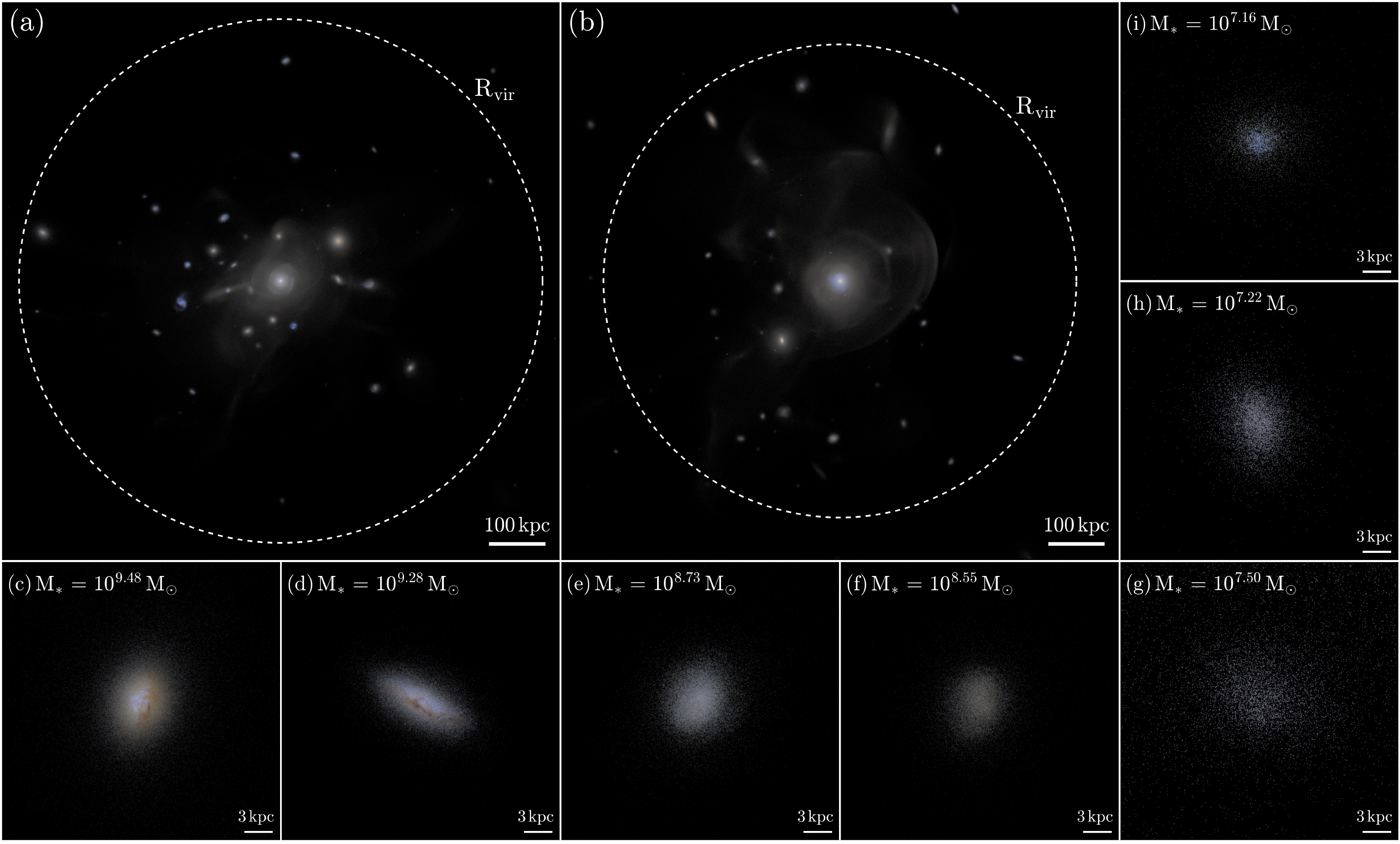}
\caption{Composite images of the two group halos (Panel-(a) and (b)) and their seven satellite galaxies randomly chosen (Panel-(c) to (i)) at $z = 0.171$.
The virial radius of each group halo is shown with the dashed circles in Panels (a) and (b).
Blue, green, and red colors are corresponding to Sloan Digital Sky Survey $g$-, $r$-, and $i$-band fluxes, respectively.}
\label{fig:fig gview}
\end{figure*}

The two most massive halos in the \NH\ volume, with virial masses\footnote{Enclosed mass of a region of which mean density is 200 times of the mean background density} of $10^{12.91}\,M_{\odot}$ and $10^{12.74}\,M_{\odot}$, are chosen as group-sized halos.
The two systems are at the limits of the typical mass range for group halos, making them somewhat less representative of group environments.
However, they serve as suitable samples for detailed investigations into the physical phenomena arising in low-mass group halos.
Throughout the study, the two groups will be referred to as Group1 and Group2, respectively.
A total of 134 galaxies ($M_{*} > 10^7\,M_{\odot}$) within three virial radii of each group are selected as sample galaxies.
The mass cut is set to ensure that the sample galaxies contain at least 1000 stellar particles.
41 of them are ``first infallers'' that have never reached their host group (see \citealp{Rhee17} for the definition of first infallers).
Since we are primarily interested in the environmental effects, we exclude them from our analysis.
12 galaxies are further removed due to their short merger tree ($<1\, {\rm Gyr}$): they have a broken merger tree or are newly detected, both of which are unsuited for our investigation.
Among the remaining 81 galaxies, two are the brightest group galaxies (BGGs), therefore Group1 and Group2 have 36 and 43 satellite galaxies, respectively, as a result.
In addition, the central positioning of the two group halos within the zoom-in region guarantees that satellite galaxies are not contaminated by low-resolution DM particles flowing into the zoom-in area from external regions.
We have confirmed this lack of contamination through our analysis.
Figure \ref{fig:fig gview} shows the composite images of the two group halos (Panel-(a) and (b)) and their satellite galaxies randomly chosen at $z = 0.171$ (Panel-(c) to (i)).
The blue, green, and red colors on each image correspond to Sloan Digital Sky Survey $g$-, $r$-, and $i$-band fluxes, respectively.
The mock images are made using SKIRT, a radiative transfer pipeline \citep[][]{BC15, CB20}.
The detailed process of image generating is described in Section 2.2 of \cite{Jang23}.

\subsection{Properties of satellite galaxies}
\label{sec:Data-Prop}
In this paper, we define the stellar mass of a galaxy as the total mass of the member stellar particles based on the 6D linking length analysis.
Similarly, we measure the SFR of galaxies based on the total mass of stellar particles that have been formed within the past $100\,{\rm Myr}$ so that our results are directly compared to observational SFR measurements.
The quenching of galaxies is determined by the birth rate parameter, $b = {\rm SFR} / M_{\rm gal} \times t_{\rm universe}$ where $t_{\rm universe}(z)$ is the age of the universe at a given redshift, which has been commonly used in various studies to identify quenched galaxies at different redshifts \citep[][]{Franx08, Lotz19, Park22}.
In this study, quenched galaxies are those with $b < 0.1$ during the last $500 \, {\rm Myr}$ consecutively, which roughly corresponds to a specific SFR of $8.59 \times 10^{-12} {\rm yr}^{-1}$ at $z = 0.171$.

To determine the gas mass of each galaxy, we consider the boundness and phase of the gas cells surrounding the galaxy.
Initially, all mass elements (DM, stars, and gas cells) within 5 $R_{\rm eff}$ of a galaxy are selected and the corresponding gravitational potential ($\Phi$) is computed.
The choice of 5 $R_{\rm eff}$ radius is empirically selected to adequately encompass the extended potential of a galaxy, avoiding the contribution from matter in neighboring galaxies to the potential.
For this computation, gas cells are treated as particles centered in their respective cells.
Cells with negative total energy, $U + K + \Phi < 0$, where $U$ and $K$ represent the internal and kinetic energy of gas cells, respectively, are considered gravitationally bound.
We further consider different phases of gas.
Cold gas cells are identified as cells with low temperatures at a given density, following the criteria proposed by \cite{Torrey12}: $\log{(T / [{\rm K}])} < 6 + 0.25 \log{(\rho / 10^{10} [M_{\odot} h^2 \, {\rm kpc}^{-3}])}$, which yields cold gas cells with temperatures between 100 and 10,000 K.

Then, we define the ISM gas mass of galaxies as the total mass of cold and bound gas cells inside $R_{\rm eff}$ of galaxies.
In addition, the SF in the \NH\ simulation occurs in gas cells with $n_{\textsc H} > 10\,{\rm cm^{-3}}$, where the SFE depends on the turbulent Mach number and the virial parameter of surrounding gas cells \citep[][]{Kimm17, Dubois21}.
Similarly, we define the dense ISM gas mass of galaxies as the total mass of cold, bound, and dense ($n_{\textsc H} > 10\,{\rm cm^{-3}}$) gas cells inside $R_{\rm eff}$, which are expected to trace \HH\ gas clump in galaxies.

The unbound gas cells are categorized into two groups: outflowing gas from the host galaxy or surrounding intragroup medium (IGM) gas.
The outflowing gas is likely a result of galactic outflow events such as SN or AGN feedback.
We define outflowing gas as gas cells with higher metallicity (than surroundings), divergent velocity vectors ($\vec{r} \cdot \vec{v} > 0$), and high temperature ($>\,10^{7}\,{\rm K}$), where $\vec{r}$ and $\vec{v}$ are the position and velocity vectors of a gas cell with respect to the galaxy center, respectively.
We use the metallicity ($Z$) threshold of $Z > \bar{Z} - \delta{Z}$,
where $\bar{Z}$ and $\delta{Z}$ are the mass-weighted mean and standard deviation of the metallicity of ISM gas cells at a given distance, respectively.
The residual unbound gas cells around galaxies are classified as IGM.
The summary of the definitions is given in Table \ref{tab:table1}.

\begin{table}
    \begin{center}
    \caption{Summary of thresholds for different gas components}
    \begin{tabular}{cc}
    \hline \hline
        Type & Conditions \\
        \hline
        ISM & bound\tablenotemark{\it a} and cold\tablenotemark{\it b} inside $R_{\rm eff}$ \\
        dense ISM & bound, cold, and dense\tablenotemark{\it c} inside $R_{\rm eff}$ \\
        outflowing & unbound, hot\tablenotemark{\it d}, metal-rich\tablenotemark{\it e} and divergent\tablenotemark{\it f} \\
        IGM & residual unbound gas \\
    \hline
    \end{tabular}
    \end{center}
    {\bf ~~Notes.}
    See the texts for the detailed definitions of the physical quantities.
    \tablenotetext{a}{$U+K+\Phi <0$}
    \tablenotetext{b}{$\log{(T / [{\rm K}])} < 6 + 0.25 \log{(\rho / 10^{10} [M_{\odot} h^2 \, {\rm kpc}^{-3}])}$}
    \tablenotetext{c}{$n_{\rm H} > 10\,{\rm cm}^{-3}$}
    \tablenotetext{d}{$T > 10^{7} {\rm K}$}
    \tablenotetext{e}{$Z > \bar{Z} - \delta Z$, where $\bar{Z}$ and $\delta Z$ are the mass-weighted mean and standard deviation of the ISM metallicity at the same distance, respectively.}
    \tablenotetext{f}{$\vec{r}\cdot\vec{v} >0$, where $\vec{r}$ and $\vec{v}$ are positional and velocity vectors of gas cells with respect to the galaxy center.}
    \label{tab:table1}
\end{table}


\section{Results}
\subsection{Quenched population of the \NH\ group galaxies}

\begin{figure*}
\centering
\includegraphics[width=0.9\textwidth]{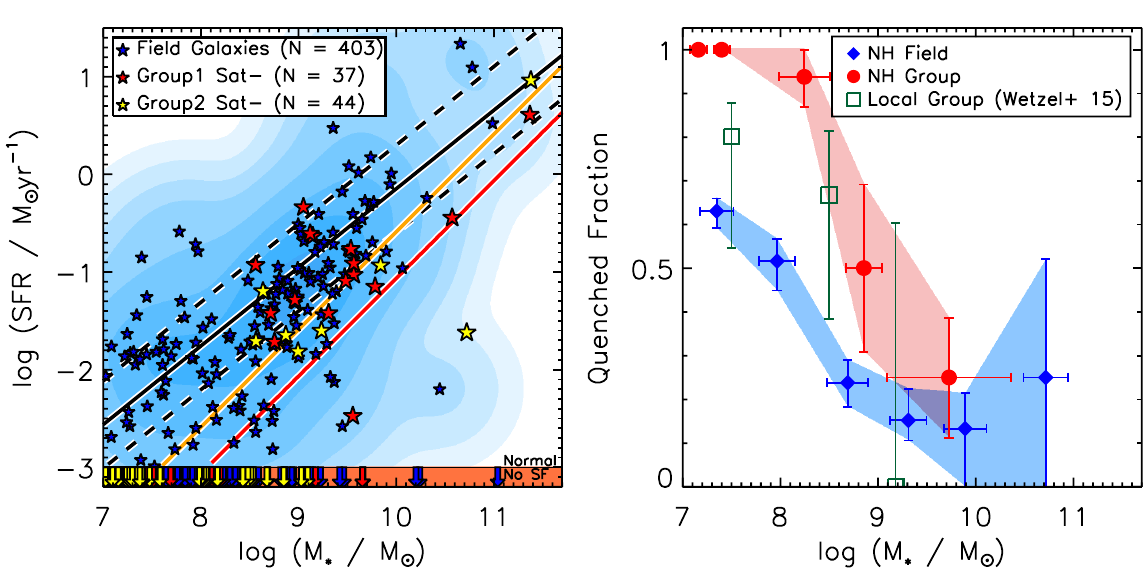}
\caption{
Main sequence relation of the \NH\ galaxies (left) and the quenched fraction of the galaxies as a function of stellar mass (right).
On the left panel, blue, red, and yellow symbols are the main sequence of the \NH\ field galaxies, satellite galaxies in Group1 and Group2, respectively.
The background blue contours represent the number density of the field sample, and the black solid and dashed lines are the linear fitting curve of the main sequence galaxies ($b > 0.3$) and its $1\sigma$ error, respectively.
The orange and red solid lines correspond to $b=0.3$ and $b=0.1$, respectively, where quenched galaxies are defined as those with $b<0.1$ for the last $500\,{\rm Myr}$.
Galaxies with no SFR are drawn with downward arrows.
On the right panel, the quenched fraction and the bootstrapping error are drawn at a fixed stellar mass bin.
Blue and red symbols are quenched fractions with the field and group galaxies, respectively.
The quenched fraction of the Local Group satellite galaxies is shown with the green open squares.
Group galaxies have a higher quenched fraction than field and Local group galaxies at all mass ranges.
}
\label{fig:fig ms}
\end{figure*}

The primary goal of this study is to identify the environmental effects on satellite galaxies in group halos.
In high-mass groups or clusters, many investigations have demonstrated a preference for passive galaxies in dense regions compared to their field counterparts \citep[e.g.,][]{Balogh04, Peng10, Wetzel13}.
The \NH\ group halos, on the other hand, are certainly low in virial mass, which may have less or no environmental effects.
Therefore, they are likely to show a different trend from what we understand with massive groups and clusters \citep[e.g.,][]{Fillingham15, Baxter21}.

To examine the quenching of SF in the satellite galaxies, we begin by plotting the relation between $M_{*}$ and SFR, main sequence relation, of field and group galaxies in the \NH\ simulation (see the left panel of Figure \ref{fig:fig ms}).
The field sample (blue stars) consists of galaxies located beyond three virial radii of any halos with $M_{\rm vir} > 10^{11} \,M_{\odot}$.
Their number density is illustrated by the background blue contours.
We fit a linear curve to the field sample with $b > 0.3$, represented by the black solid line, and include the $1\sigma$ errors with black dashed lines.
We also display the group galaxies with red (Group1) and yellow (Group2) symbols, with downward arrows indicating galaxies with no SF activity.
To facilitate identifying quenched and star-forming galaxies, we show the lines with $b=0.1$ and $b=0.3$ as red and orange solid lines, respectively.

\begin{figure*}
\centering
\includegraphics[width=0.9\textwidth]{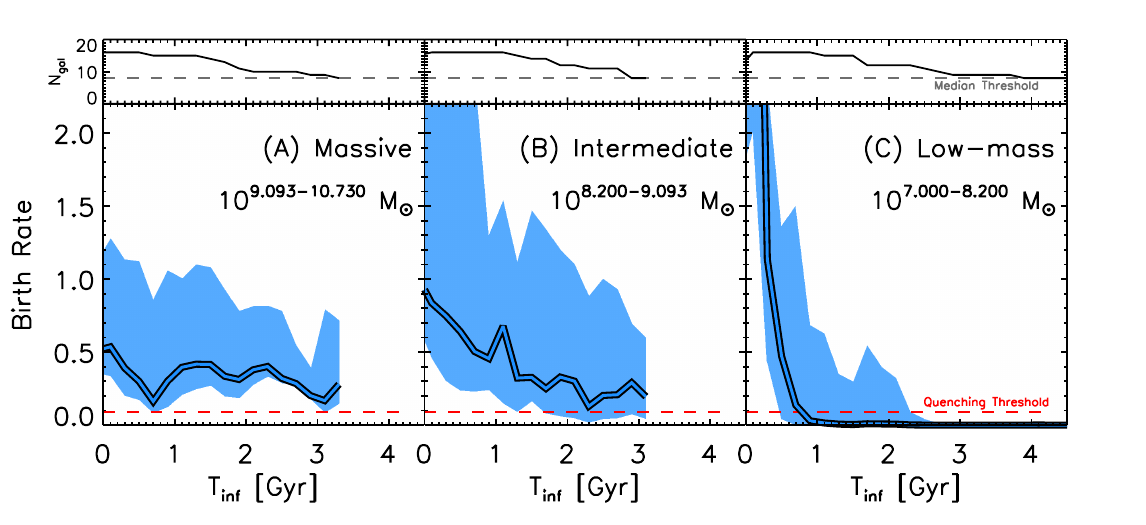}
\caption{Star formation histories of the sample galaxies as a function of time since first infall of galaxies.
The star formation rate of galaxies is replaced with the birth rate of galaxies.
The blue solid lines and shaded areas correspond to the median and the $1^{\rm st}$ and $3^{\rm rd}$ quartiles of the distribution of individual lines, respectively.
The red dashed line shows the quenching threshold (birth rate of 0.1) used in this study.
Massive and intermediate-mass galaxies show complex star formation history patterns.
These galaxies experience quenching shortly after their infall, and a significant number of them undergo subsequent rejuvenation.
Low-mass galaxies are also quenched after infall and only a few of them have rejuvenated.
Upper panels show the number of galaxies involved with the median calculation (black solid line).
The grey dashed line indicates the minimum number of galaxies (8) for the median calculation.
}
\label{fig:fig nqsfh}
\end{figure*}

The main sequence of the \NH\ field galaxies is evident, as reported in \cite{Dubois21}.
However, the group galaxy sample displays a different characteristic: most of them with $M_{*} < 10^{8.5} \, M_{\odot}$ are quenched.\footnote{$b<0.1$ for the last 500 ${\rm Myr}$}
The presence of ``star-forming'' galaxies in the same mass range among the field galaxies suggests that low-mass galaxies have been affected by environmental effects.
In contrast, around $68\%$ of the group galaxies with $M_{\rm *} > 10^{8.5} \, M_{\odot}$ are either star-forming ($b>0.3$) or in the green valley ($0.3>b>0.1$).
This suggests that environmental quenching, if it exists in group halos, is not as effective for massive galaxies as it is for low-mass galaxies.

On the right panel in Figure \ref{fig:fig ms}, the quenched fractions of the field (blue) and group (red) galaxies are presented as a function of stellar mass.
The quenched fraction at each stellar mass bin and its $1\sigma$ error (shaded region) are obtained by bootstrapping sampling.
The quenched fractions of the LG satellites \citep[][]{Wetzel15} are also shown with open green squares.

It is evident that the quenched fraction of group galaxies is higher than that of field galaxies across all mass ranges.
This implies that environmental quenching is prevalent in the groups across a wide mass range.
A decreasing trend with increasing stellar mass of quenched fraction is clear, which is consistent with earlier findings from simulations \citep[][]{Akins21, Karunakaran21, Samuel22}.

The \NH\ group galaxies show higher quenched fractions compared to the field galaxies.
However, 31 out of the 79 galaxies were already quenched prior to their accretion into the host groups.
Hence, these 31 quenched galaxies are unsuited to study group environmental effects, and we have decided to exclude them from our sample.
Roughly half of them are formed with a starburst phase and then subsequently quenched due to subsequent strong SN feedback.
The other ex-situ quenched galaxies slowly enter a passive state by failing to form dense ISM gas where SF can take place.
The remaining 48 satellite galaxies constitute our main sample.


\subsection{Features of quenched galaxies}

\subsubsection{Star formation history}

We begin by investigating the star formation history (SFH) of the 48 galaxies to see if there is any specific pattern of SF quenching inside group halos.
For example, cluster galaxies are known to share a common SF quenching pattern \citep[][]{Wetzel13}, characterized by a delay in the cessation of SF following the accretion to clusters.

As indicated by the mass-dependent quenching pattern in the right panel of Figure \ref{fig:fig ms}, we divide the 48 galaxies into three groups by their stellar mass: massive ($10^{9.093-10.730}\, M_{\odot}$), intermediate-mass ($10^{8.200-9.093}\, M_{\odot}$), and low-mass ($10^{7.000-8.200}\, M_{\odot}$).
These mass bins are determined using equal-number statistics of the galaxies.
Figure \ref{fig:fig nqsfh} shows the SFH of these galaxies after their first infall\footnote{The first arrival at one virial radius} into the host group.
As introduced in Section \ref{sec:Data-Prop}, we employ the birth rate parameter as an SFR indicator so as to mitigate variations in the SFR of galaxies with different stellar masses and different redshifts.
The median of the evolution, shown with the blue solid lines, is computed when there are more than eight galaxies in each time bin ($0.2 \,{\rm Gyr}$), with the shaded area indicating the $1^{\rm st}$ and $3^{\rm rd}$ quartiles.
Thus, the median evolution represents systematic changes in the birth rate of individual galaxies at each time bin.
The number of galaxies used to calculate the median in each time bin is shown in the upper panels with the black solid line.
Time is normalized by the infall epoch of the galaxies, with 0 representing the infall moment.
The quenching threshold ($b = 0.1$) is shown with the red dashed line.


Contrary to cluster galaxies, which are generally expected to be quenched soon after one orbit \citep[e.g.,][]{Lotz19, Rhee20}, the group satellite galaxies have a complex SFH similar to what has been observed from the LG dwarf galaxies \citep[][]{Weisz14}.
Massive and intermediate-mass galaxies maintain their SF activity for $T_{\rm inf} > 3\,{\rm Gyr}$.
This timescale significantly exceeds the crossing timescale of galaxies, indicating that galaxies in the low-mass groups continue forming stars during multiple orbital motions.
In addition, massive galaxies show an initial decrease in the birth rate after the infall, followed by subsequent periods of SF rejuvenation, although the statistical significance of this trend is somewhat limited.
These trends will be further investigated in Section \ref{sec:Res-orbit}.
Thus group environmental effects are working not as an abrupt cessation of SF but rather as a modulation of SF activity over time.
On the other hand, low-mass galaxies are rapidly quenched shortly after infall.
This is perhaps due to their shallow potential well, resulting in limited ability to sustain their SF activity.

\subsubsection{The origin of variation in SF}
\label{sec:result-SF}

\begin{figure*}
\centering
\includegraphics[width=0.9\textwidth]{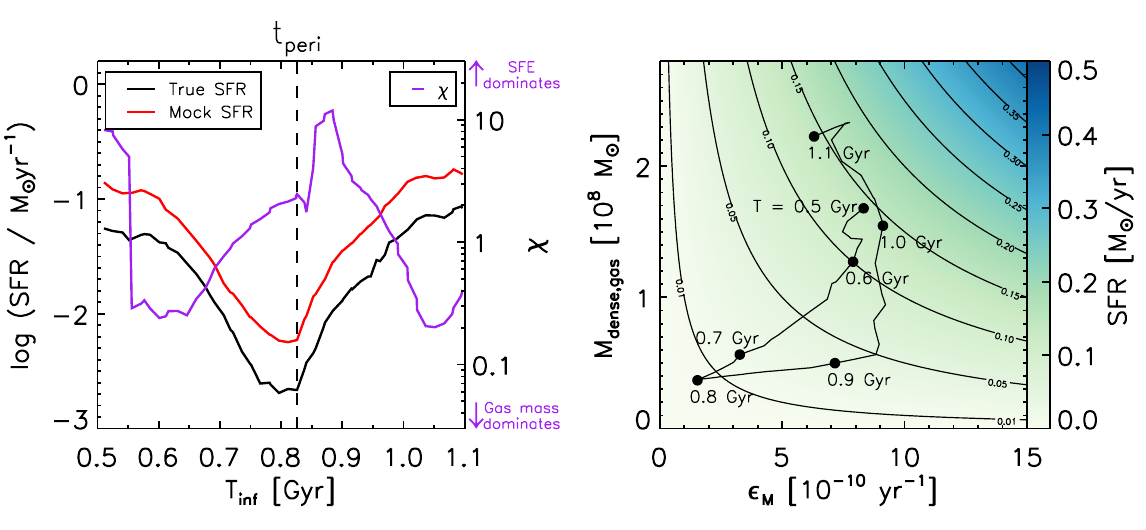}
\caption{Left panel: The SFR and $\chi$ value evolution of one galaxy around the pericenter passage.
True SFR (black solid line) is measured with newly formed stars (see Section \ref{sec:Data} for the exact definition) and mock SFR (red solid line) is the SFR predicted by Equation \ref{eq:eq SF}.
The purple solid line indicates $\chi$ value evolution, which represents the relative contribution of star formation efficiency and gas mass to the SFR.
The mock SFR has a good predictive power of the true SFR, and $\chi$ quantitatively shows the origin of SFR variation.
Right panel: Evolution of the gas mass and star formation efficiency of the same galaxy to the left panel over the same $T_{\rm inf}$ range.
The multiplication of these two quantities corresponds to the mock SFR of this galaxy which is shown with the background colors and contours.}
\label{fig:fig nqsfphase}
\end{figure*}



To check out the origin of changes in SFR seen in Figure \ref{fig:fig nqsfh}, we now aim to establish a connection between the SFR of a galaxy and the galactic bulk properties.
In the \NH\ simulation, SF takes place in gas cells with $n_{\textsc H} > 10\,{\rm cm^{-3}}$ following a Schmidt law \citep[][]{Dubois21}:
\begin{equation}
\dot{\rho}_{*} = \frac{\varepsilon_{*}}{t_{\rm ff}} \rho_{\rm gas},
\label{eq:eq SLaw}
\end{equation}
where $\rho_{*}$ is the mass density of newly formed stars, $\rho_{\rm gas}$ is the gas mass density of a cell, and $t_{\rm ff} = \sqrt{3\pi / (32G\rho_{\rm gas}})$ is the free-fall time of the gas cell, $G$ being the gravitational constant.
The efficiency parameter, $\varepsilon_{*}$ is determined by the gravo-turbulent SF model \citep[][]{Kimm17, Dubois21}, and it depends on the turbulent Mach number, $\mathcal{M}$, and the virial parameter, $\alpha = |2E_{\rm kin} / E_{\rm grav}|$ of the gas cell.
Here, $E_{\rm kin}$ and $E_{\rm grav}$ are the kinetic and local gravitational potential energies, respectively.
In general, $\varepsilon_{\rm *}$ is increasing with an increasing Mach number and a decreasing virial parameter (see Figure 1 in \citealp{FK12}).
The total SFR of a galaxy ($\phi$) is then the total mass of newly formed stars ($\dot{M}_{*}$) per a time range ($\Delta T$):
\begin{equation}
\phi(t) = \frac{1}{\Delta T} \int^{t}_{t-\Delta T} dt \, \dot{M}_{*}.
\end{equation}
The time differential of the mass of new stars now can be estimated with volume integration of Equation \ref{eq:eq SLaw} over that galaxy:
\begin{equation}
\dot{M}_{*} = \int dV \rho_{\rm gas} \frac{\varepsilon_{*}}{t_{\rm ff}}.
\label{eq:eq Mstar}
\end{equation}
The Equation \ref{eq:eq Mstar} can be re-written as follows:
\begin{equation}
\begin{aligned}
    \dot{M}_{*} &= \frac{\displaystyle \int dV \rho_{\rm gas} \frac{\displaystyle \varepsilon_{*}}{\displaystyle t_{\rm ff}}}{\displaystyle \int dV \rho_{\rm gas}} \int dV \rho_{\rm gas} \\
    &= \epsilon_{M} \int dV \rho_{\rm gas},
\end{aligned}
\end{equation}
where $\epsilon_{M}$ is the mass-weighted average of $\frac{\varepsilon_{*}}{t_{\rm ff}}$.
Here, we define the volume of a galaxy as a sphere with a radius of $R_{\rm eff}$ centered at the galaxy center.
Then, $\int dV \rho_{\rm gas}$ is equal to the dense ISM gas mass ($M_{\rm dense, gas}$) of a galaxy\footnote{Most of the gas cells with $n_{\rm H} > 10\,{\rm cm}^{-3}$ inside $R_{\rm eff}$ satisfy the other conditions for dense ISM.} defined in Section \ref{sec:Data-Prop}.
Eventually, the SFR of a galaxy predicted from the SF model in \NH\ is
\begin{equation}
\begin{aligned}
\phi(t) &= \frac{1}{\Delta T} \int^{t}_{t - \Delta T} dt \, \epsilon_{M} M_{\rm dense,gas} \\
    &=  \epsilon_{M} M_{\rm dense, gas},
\end{aligned}
\label{eq:eq SF}
\end{equation}
where the last correspondence has the assumption that both $\epsilon_{M}$ and $M_{\rm dense,gas}$ remain constant over a short time range ($\Delta T = 100 {\rm Myr}$).
Indeed, we use the time average of both parameters over $\Delta T$.
Therefore, the SFR of galaxies can be predicted by the two bulk parameters of galaxies, $\epsilon_{M}$ and $M_{\rm dense, gas}$, which are measurable through resolved the Kennicutt-Schmidt relation for example.
However, it is important to note that this simplified model may introduce some systematic uncertainties, such as mass loss during SF (mass changes of newly formed stars) and variations in free-fall timescales for different gas cells (violation of the assumptions in Equation \ref{eq:eq SF}).

The left panel of Figure \ref{fig:fig nqsfphase} shows the validation of our SFR prediction using Equation \ref{eq:eq SF}.
In this panel, we compare the actual SFR evolution of one galaxy around its pericenter passage (shown as a black solid line) with the predicted SFR with Equation \ref{eq:eq SF} (red solid line).
The mock SFR reasonably reproduces the original SFR values, generally within a few factors, although we acknowledge the presence of some systematic bias in our SFR prediction mentioned previously.
We confirm that the mock SFR estimation well imitates the original values for other sample galaxies.
Hence, the predicting power of $\epsilon_{M}$ and $M_{\rm dense, gas}$ allows us to analyze variations in the SFR of galaxies by considering changes in both bulk parameters.

For example, the right panel illustrates the evolution of $\epsilon_{M}$ and $M_{\rm dense, gas}$ of the same galaxy on the $\epsilon_{M}$-$M_{\rm dense, gas}$ plane, where the product of the two physical quantities represents the mock SFR of the galaxy.
The background colors and contours indicate the corresponding value of SFR ($\epsilon_{M} \times M_{\rm dense, gas}$).
In this particular galaxy, there is an initial decline in SFR during the period of $T_{\rm inf} = 0.5 - 0.8 \, {\rm Gyr}$, primarily driven by the reductions in both $\epsilon_{M}$ and $M_{\rm dense, gas}$.
Subsequently, the SFR increases mainly due to the rise in the $\epsilon_{M}$ during $T_{\rm inf} = 0.8 - 0.9 \, {\rm Gyr}$, followed by an increase in gas mass at $T_{\rm inf} \gtrsim 0.9 \, {\rm Gyr}$.

To quantitatively estimate the relative contributions of SFE and gas mass to the SFR values, we introduce another parameter.
Assuming that $\phi = \epsilon_{M} M_{\rm dense, gas}$ accurately captures the true SFR values, the time derivative of the SFR can be expressed analytically as follows:
\begin{equation}
\begin{aligned}
\frac{d \phi}{dt} &= \frac{\partial \phi}{\partial \epsilon_{M}} \frac{d \epsilon_{M}}{dt} + \frac{\partial \phi}{\partial M_{\rm dense, gas}} \frac{d M_{\rm dense, gas}}{dt} \\
    &= M_{\rm dense, gas} \frac{d \epsilon_{M}}{dt} + \epsilon_{M}\frac{d M_{\rm dense, gas}}{dt}.
\end{aligned}
\end{equation}
The former and latter terms represent the time derivative of SFR resulting from variations in $\epsilon_{M}$ and $M_{\rm dense, gas}$, respectively.
Then, the magnitude of the fraction of these terms, 
\begin{equation}
\chi \equiv \frac{ | M_{\rm dense, gas} \displaystyle \frac{d \epsilon_{M}}{dt} | }{| \epsilon_{M} \displaystyle \frac{d M_{\rm dense, gas}}{dt} |}
\end{equation}
provides an estimate of the contribution from each term.
A higher value of $\chi$ ($>1$) indicates a greater impact of SFE on the changes in SFR.

In the left panel of Figure \ref{fig:fig nqsfphase}, the advantage of employing $\chi$ to understand variations in SFR becomes apparent.
During the declining phase of SFR ($T_{\rm inf} = 0.5 - 0.6 \, {\rm Gyr}$), the primary driving factor is the decrease in $M_{\rm dense,gas}$ (see the right panel).
Consequently, the values of $\chi$ fall below 1.
Subsequently, in the interval of $T_{\rm inf}=0.6-0.8$, both $\epsilon_{M}$ and $M_{\rm dense,gas}$ are decreasing, and $\chi$ increases, indicating that the contribution of SFE becomes more pronounced during this period.
Then, during the early rejuvenation phase ($T_{\rm inf} = 0.8 - 0.9 \, {\rm Gyr}$), the value of $\chi$ reaches its highest point, corresponding to the horizontal movement of the SFR evolution in the right panel.
Finally, during the later rejuvenation phase ($T_{\rm inf} = 0.9 - 1 \, {\rm Gyr}$), the vertical motion in the right panel aligns well with a decreasing trend of $\chi$ during the same time period, implying stronger contribution from gas mass.

Hence, we try to measure quantities within the small-scale physical areas of galaxies, confirming that their associated bulk properties have a robust predictive power for SFR and the potential to investigate the origins of the SFR variation with time.
This insight is made possible by the high-resolution characteristics of the \NH\ simulation, enabling the measurement of ISM properties at sub-hundred-parsec scales with high temporal resolution ($\sim 15\,{\rm Myr}$).
This approach may represent an early attempt to study the environment-driven response of galaxies down to their small-scale ISM dimensions within a cosmological simulation.


\subsubsection{Orbit-related quenching}
\label{sec:Res-orbit}

Cluster galaxies have been consistently reported to have a quenching pattern associated with their orbital motion \citep[][]{Mock14, Foltz18, Lotz19, Rhee20, Oman21}.
This pattern suggests that galaxies undergo a suppression of SF at their pericenter passages. 
The predominant explanation for this phenomenon is that the strong ram-pressure stripping leads to the removal of the ISM and subsequent SF quenching in the vicinity of cluster centers.
Remarkably, the SFR evolution of the galaxy shown in Figure \ref{fig:fig nqsfphase} mirrors this previously identified trend.
This finding is particularly intriguing as it suggests that even in low-mass groups, where ram pressure is less pronounced, there is evidence of an orbit-related quenching feature similar to that witnessed in cluster environments.

\begin{figure*}
\centering
\includegraphics[width=0.75\textwidth]{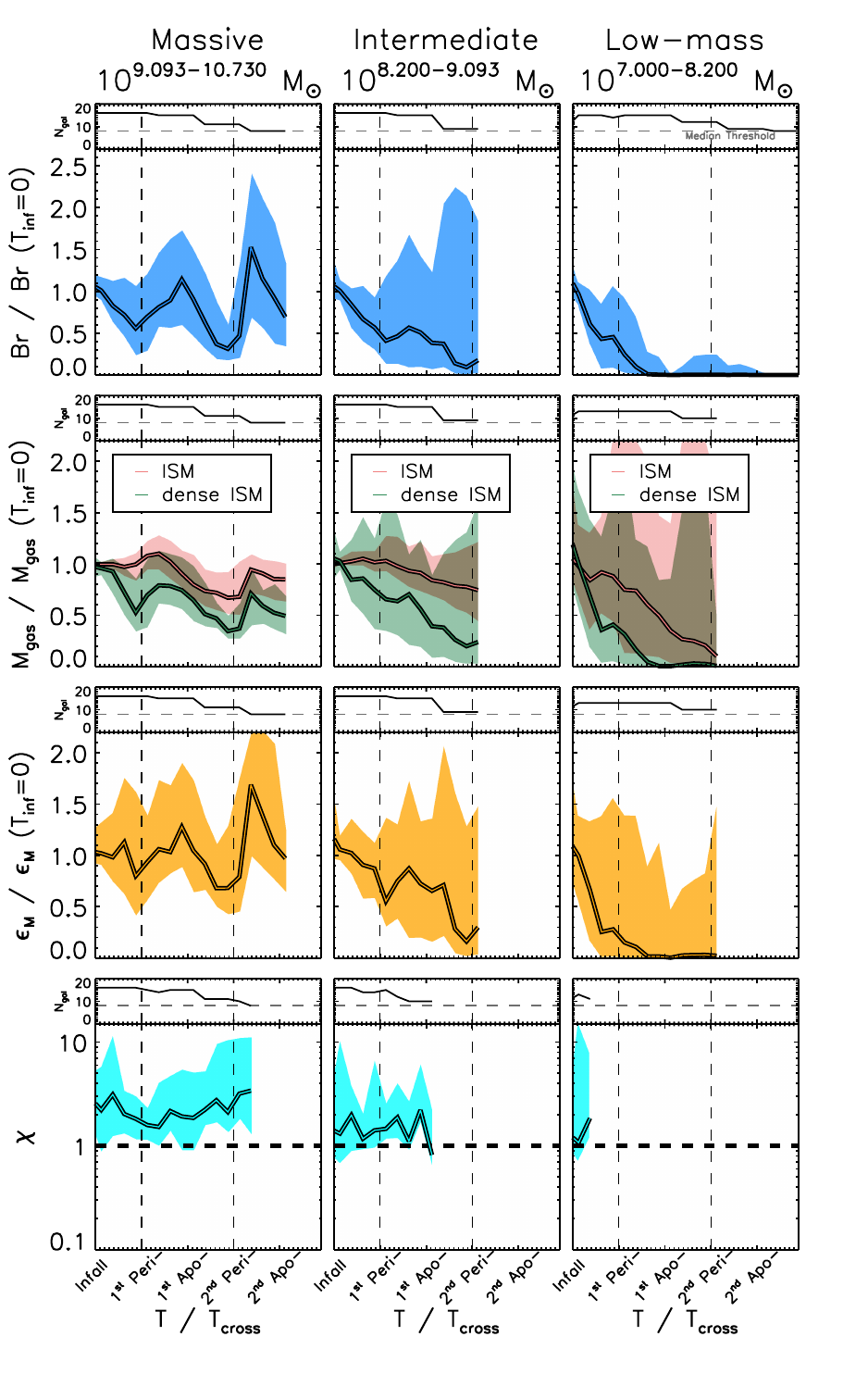}
\caption{
Birth rate (Br), gas masses, $\epsilon_{M}$, and $\chi$ value evolution of the sample galaxies.
The first, second, and third columns are evolution with massive, intermediate-mass, and low-mass galaxies, respectively.
All panels have the same format as Figure \ref{fig:fig nqsfh}.
The time since first infall of galaxies is normalized by their crossing times so that the galaxies will have the same pericenter passages and apocenter passage times (see the text for details).
Birth rate, gas masses, and $\epsilon_{M}$ are normalized by the values of galaxies at their first pericenter to reduce galaxy-to-galaxy variations.
}
\label{fig:fig nqorbit}
\end{figure*}

However, the median of SFHs in Figure \ref{fig:fig nqsfh} implies complex variations in the individual SFHs of galaxies.
To mitigate galaxy-to-galaxy variations, which stem from different orbital times, let us standardize the infall time of each galaxy relative to its crossing time.
This enables us to investigate the evolution of SFR concerning their pericenter and apocenter passages.
First, we measure the pericenter and apocenter passages for each galaxy.
Then, we can define a series of passage times: ${t_{\rm 0} < t_{\rm p,1} < t_{\rm a,2} < t_{\rm p,3} < t_{\rm a,4}} < \cdots$, where $t_{\rm 0}$ is the infall moment and $t_{\rm p,n}$ and $t_{\rm a,m}$ are the pericenter and apocenter passage times, respectively.
We employ a simple linear transformation of galaxy infall times to derive the normalized time:
\begin{equation}
    \begin{aligned}
    \frac{T}{T_{\rm cross}} = \begin{dcases}
        \frac{T_{\rm inf} - t_{\rm p,n}}{t_{\rm a,n+1} - t_{\rm p,n}} + n \quad \text{if } t_{\rm p,n} < T_{\rm inf} < t_{\rm a,n+1} \\
        \frac{T_{\rm inf} - t_{\rm a,m}}{t_{\rm p,m+1} - t_{\rm a,m}} + m \quad \text{if } t_{\rm a,m} < T_{\rm inf} < t_{\rm p,m+1}.
    \end{dcases}
\end{aligned}
\end{equation}
For example, galaxies with $0 < T / T_{\rm cross} < 1$ correspond to those that are first infalling, while galaxies with $1 < T / T_{\rm cross} < 2$ are moving away after their first pericenter passage.

Figure \ref{fig:fig nqorbit} shows the evolution of the birth rate (Br), the ISM and dense ISM gas masses, $\epsilon_{M}$, and $\chi$\footnote{We note that $\chi$ is not measured for galaxies with very low SFR ($<10^{-3} M_{\odot} {\rm yr}^{-1}$) due to uncertainties arising from a low number of star-forming gas cells.} of galaxies as a function of $T / T_{\rm cross}$.
To minimize variations among galaxies, the birth rate, gas masses, and $\epsilon_{M}$ are normalized to the values at $T_{\rm inf} = 0$.
The median evolution of each property is computed similarly as in Figure \ref{fig:fig nqsfh}.
The three columns, from left to right, represent the evolution of galaxies in massive, intermediate-mass, and low-mass bins, respectively.
The purpose of this figure is to investigate whether the SFH and related properties of galaxies are associated with the orbital motions in low-mass groups.

In contrast to the intricate variations in the SFH shown in Figure \ref{fig:fig nqsfh}, the evolution of the birth rate with normalized time shows a clear connection with orbital motions.
The median curves of galaxies display decreasing birth rate trends from infall to the first pericenter passages.
While both massive and intermediate-mass galaxies maintain extended periods of SF, it appears that only massive galaxies have multiple rejuvenation phases following their pericenter passages, resulting in periodic SFH.
Intermediate-mass galaxies, on the other hand, have substantial SFR fluctuations after the first pericenter passages, eventually becoming quenched at the second pericenter passage.
Low-mass galaxies have a steadily decreasing birth rate, which is totally quenched after the first pericenter passage, likely due to their shallower gravitational potential.
Thus, quenching timescales, if defined as the elapsed time from the infall to the final quenching moment, show a mass trend, increasing with increasing stellar mass.
This may indicate that only massive galaxies can sustain their SF phase beyond the pericenter passage, while others fail to do so.
It differs significantly from the cluster case in that some galaxies display periodic SFH within groups.

As introduced in Figure \ref{fig:fig nqsfphase}, dense gas mass and $\epsilon_{M}$ are proved to be useful for investigating the origin of SF variation.
In the case of massive galaxies, the change in ISM gas mass is relatively modest (red solid line in the second row and the first column).
This implies that these galaxies may experience mechanisms that lead to gas loss, but significant ISM stripping does not occur over several orbital times.
Intriguingly, the dense ISM gas mass of the massive galaxies has a similar pattern in their evolution to the birth rate evolution, having a valley at the pericenter passages.
Hence, during the orbital motions of massive galaxies, their gas components are affected primarily by changing their phase rather than being stripped.
Considering that they still retain more than $\sim 50 \%$ of their initial gas after the second pericenter passage, they may still have a sufficient amount of gas for SF in the future.

Similarly, $\epsilon_{M}$ values for the massive galaxies (the third row and the first column) show a similar trend to the birth rate, featuring local minima at pericenter passages.
The oscillating behavior of SFE and dense gas closely resembles the pattern observed in the birth rate evolution.
This suggests a collaborative effort between these factors in quenching and rejuvenation of SF.
Their relative importance can be quantitatively estimated by examining the evolution of $\chi$ in the bottom row.
Throughout all epochs, $\chi$ remains consistently greater than 1 for massive galaxies, suggesting that SFE contributes more to SFR than gas mass does.
Therefore, massive galaxies have periodic SFH associated with their orbital motions, more likely driven by changes in SFE.

Intermediate-mass galaxies show somewhat different properties.
Following their infall, both ISM and dense ISM gas masses are gradually decreasing, whereas dense ISM gas mass is nearly depleted at the second pericenter passage.
This means that the lack of dense ISM gas becomes more critical for their SF quenching.
At their first infall phase, both gas mass and SFE are decreasing, corresponding to the initial decline of the birth rate at the same period.
During their orbital motions, $\chi$ values are around 1, indicating that both gas mass and SFE contribute approximately equally.
At their final quenching moment, both SFE and gas mass are notably low, suggesting that the final quenching results from reductions in both factors, with comparable contributions from each.
Therefore, the quenching of intermediate-mass galaxies is influenced by both gas mass and SFE.

Low-mass galaxies show a more pronounced evolution.
They become fully quenched following the first pericenter passage.
Simultaneously, both their dense ISM gas mass and SFE decrease, suggesting that SF quenching is likely attributed to both factors.
The $\chi$ value is temporarily around 1 but then becomes undefined due to the measurement threshold with SFR $> 10^{-3} M_{\odot} {\rm yr}^{-1}$.
Hence, it is likely that the birth rate of low-mass galaxies is initially decreasing with comparable contributions from $\epsilon_{M}$ and dense ISM gas mass, but they are eventually quenched due to deficiency of dense ISM gas mass after the first pericenter passage.
The median evolution of ISM gas mass also quickly dropped after the first pericenter passage.
This indicates that these galaxies undergo significant gas loss after the pericenter passage, likely due to ram pressure.

Therefore, the satellite galaxies show a clear behavior of SF quenching related to their orbital motion.
However, we acknowledge the potential bias introduced by the redshift evolution of the systems.
In Figure \ref{fig:fig nqorbit}, we compute the median of sample galaxies at various redshifts.
While we employed the normalized time on the axis to align galaxies at the comparable orbital stages, our analysis does not account for the redshift evolution of the group systems.
This omission could potentially bias the results discussed in this subsection, particularly if the group environmental effects vary with redshift.

To address this, we tested whether the main finding of Figure \ref{fig:fig nqorbit}, namely the orbit-related quenching feature, persists across satellite galaxies with varying time since infall.
We initially examined the evolutionary history of individual satellite galaxies and confirmed the presence of the orbit-related quenching features in these cases.
Subsequently, we investigated whether this trend is influenced by the time since infall of satellite galaxies.
The orbit-related quenching feature is consistently observed in satellite galaxies, regardless of whether they were accreted earlier (the half of galaxies with $T_{\rm inf}$ lower than their median value) or later (the other half).
However, those accreted earlier generally show a mild level of quenching, such as a higher value of birth rate at the first pericenter.
This may suggest that the environmental effects in the early stages of group halos might be less intense compared to current group systems.
Due to the small number of samples, further quantitative analysis is not feasible, and a detailed study of the redshift dependency of group environmental effects will be pursued in our future investigation.






\section{Discussion}

\subsection{Origin of changes in gas mass}


\begin{figure*}
\centering
\includegraphics[width=0.95\textwidth]{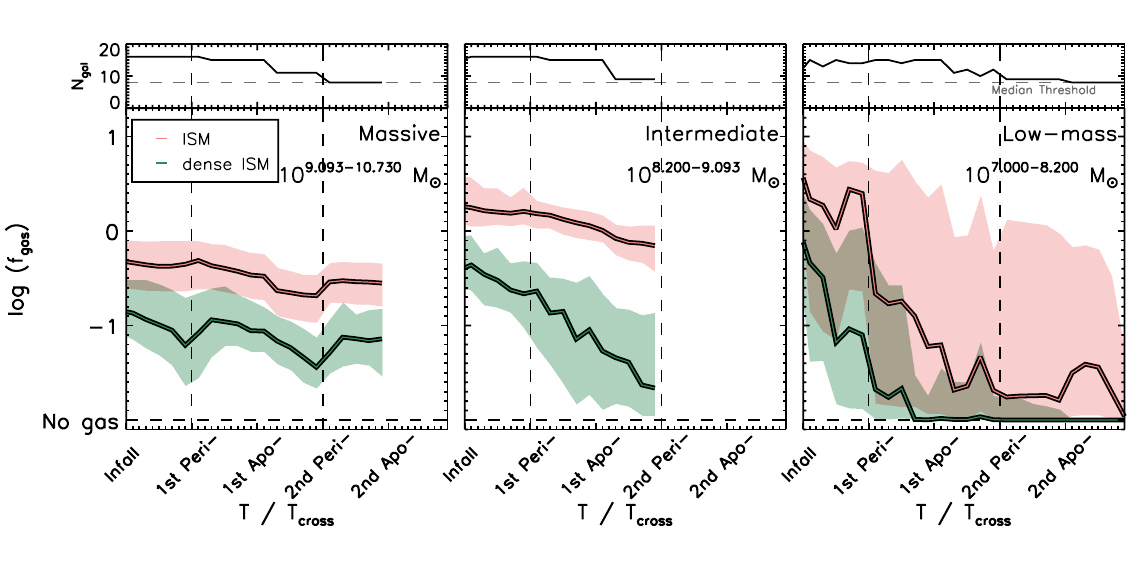}
\caption{
ISM and dense ISM gas evolution of galaxies.
The figure has the same format as the second row in Figure \ref{fig:fig nqorbit} but presents the gas fractions instead of the normalized ones.
}
\label{fig:fig nqgmass}
\end{figure*}

In Section \ref{sec:Res-orbit}, we discussed how both the decrement of dense ISM gas and a change in SFE contribute to the final SF quenching of satellite galaxies.
In the following subsections, we explore the underlying causes of changes in these two properties.

Figure \ref{fig:fig nqgmass} illustrates the evolution of gas fractions ($f_{\rm gas} = M_{\rm gas} / M_{*}$) of the ISM and dense ISM gas of galaxies.
The figure follows the same format as the second row in Figure \ref{fig:fig nqorbit}, but instead of presenting normalized values, it shows the actual gas fractions.
In the figure, we classify galaxies with $\log{f_{\rm gas}}<-2$ as gas-deficient galaxies.

The ISM gas content in massive galaxies remains relatively constant over time, indicating that these galaxies do not undergo substantial gas removal processes.
It is likely that their deeper potential wells allow them to retain their gas components even after experiencing multiple orbital cycles.
However, a different pattern emerges for the dense ISM gas.
As shown in Figure \ref{fig:fig nqorbit}, the amount of dense ISM gas diminishes at the pericenter passages and then rebounds.
This suggests that certain mechanisms at the pericenter passages affect the phase of the gas components in massive galaxies (see Section \ref{sec:Dis-SFE}).

To gain further insights, we investigate the fractional size evolution of the dense and total ISM gas ($R_{\rm dense ISM} / R_{\rm ISM}$)  in Figure \ref{fig:fig nqgmass2}.
In this context, the size of gas components is defined as their half-mass radius.
Their fractional size shows a similar periodic pattern, indicating that the size of the dense ISM gas contracts at the pericenter passage and subsequently expands.
One possible explanation is that the outer part of the dense ISM gas is disrupted at the pericenter passages due to interactions with hot IGM gas blown onto them.
Hence, massive galaxies undergo a cycle in their gas components: the dense ISM gas becomes concentrated and decreases in mass at the pericenter, and then it redistributes and increases in mass afterward.
Following this cycle in the gas phase, the SFR of massive galaxies is affected, albeit with a weaker contribution compared to SFE.

Intermediate-mass galaxies are also relatively constant in their amount of ISM gas, indicating that environmental processes to which galaxies are subject do not have a significant impact on their gas components.
In fact, only three galaxies eventually become gas-deficient at the final moment.
Considering the notable decline in gas content observed in low-mass galaxies (see the right panel of Figure \ref{fig:fig nqgmass}), a threshold for the effective gas removal process appears to exist around $10^{8}\,M_{\odot}$.
In contrast, the dense ISM gas of intermediate-mass galaxies decreases more strongly over time.
The median curve for their dense ISM gas fraction reaches $\log{f_{\rm gas}} < -1$ at around the second pericenter passages, indicating that over half of these galaxies become deficient in dense ISM gas.
This dearth of dense ISM gas is interesting, given the concurrent presence of a substantial amount of ISM gas during the same period.
This suggests that a particular physical process may obstruct the conversion from ISM gas to dense ISM gas.
Although not shown here, their fractional size of dense ISM and ISM gas indeed decrease similarly at the first pericenter, with only a few of these galaxies returning to their original size.
Thus, it is possible that intermediate-mass galaxies undergo a similar cycle to massive galaxies but only once.
Thus, likely due to their relatively shallower potential, intermediate-mass galaxies face challenges in replenishing their dense ISM gas after the first pericenter passage, resulting in a gradual decrease in their dense ISM gas content.

Low-mass galaxies show remarkable features.
The ISM gas quantity remains relatively constant from the infall to the first pericenter passage of galaxies.
At this stage, over half of the low-mass satellite galaxies are rich in gas, although two of them become gas-poor around the first pericenter passage.
This transition contributes to a downturn in the median curve at the first pericenter, where one galaxy experiences gas stripping due to interactions and another from ram pressure stripping.
Considering the abundance of gas-rich low-mass galaxies at their first infall phase, this implies that ram pressure stripping might not be the primary mechanism responsible for gas depletion in these low-mass galaxies.

Following the pericenter passages, both ISM and dense ISM gas of the low-mass galaxies show a rapid decline.
Therefore, as previously discussed, the primary driver of SF quenching of the low-mass satellites is the removal of gas.
Therefore, a stellar mass of $10^{8}\,M_{\odot}$ acts like a threshold for gas-stripping of the NH satellite galaxies, consistent with the reported range with the MW satellites \citep[e.g.,][]{Fillingham16}.
In our inspection of individual galaxies, the gas removal generally occurs concurrently with the application of strong external pressures.
This includes ram pressure stripping by the IGM, hydrodynamic interactions with nearby galaxies, and outflows from adjacent galaxies.


However, the upper quartile of the ISM gas fraction remains high even after second pericenter passages, indicating that at least $25 \%$ of low-mass galaxies involved in the median computation (the number is shown with the black-dashed line) retain a sufficient amount of ISM gas.
The Jeans length of the ISM gas in these galaxies ranges from $0.1 - 1\, {\rm kpc}$, suggesting that the \NH\ spatial resolution ($34\,{\rm pc}$ in comoving length) is adequate to resolve the clumpy structure of ISM components and mitigates the possibility of an artificial quenching in low-mass galaxies discussed in \cite{Schaye15}.
Given the sufficient amount of their ISM gas, a certain fraction of low-mass galaxies undergo quenching due to an inability to form dense ISM gas.
These galaxies might share a common origin with some of the satellite galaxies in the local volume (e.g., NGC 4163 or Sextans B), with low levels of SF activity and high gas fractions \citep[][]{GP09, Karachentsev13}.




\begin{figure}
\centering
\includegraphics[width=0.45\textwidth]{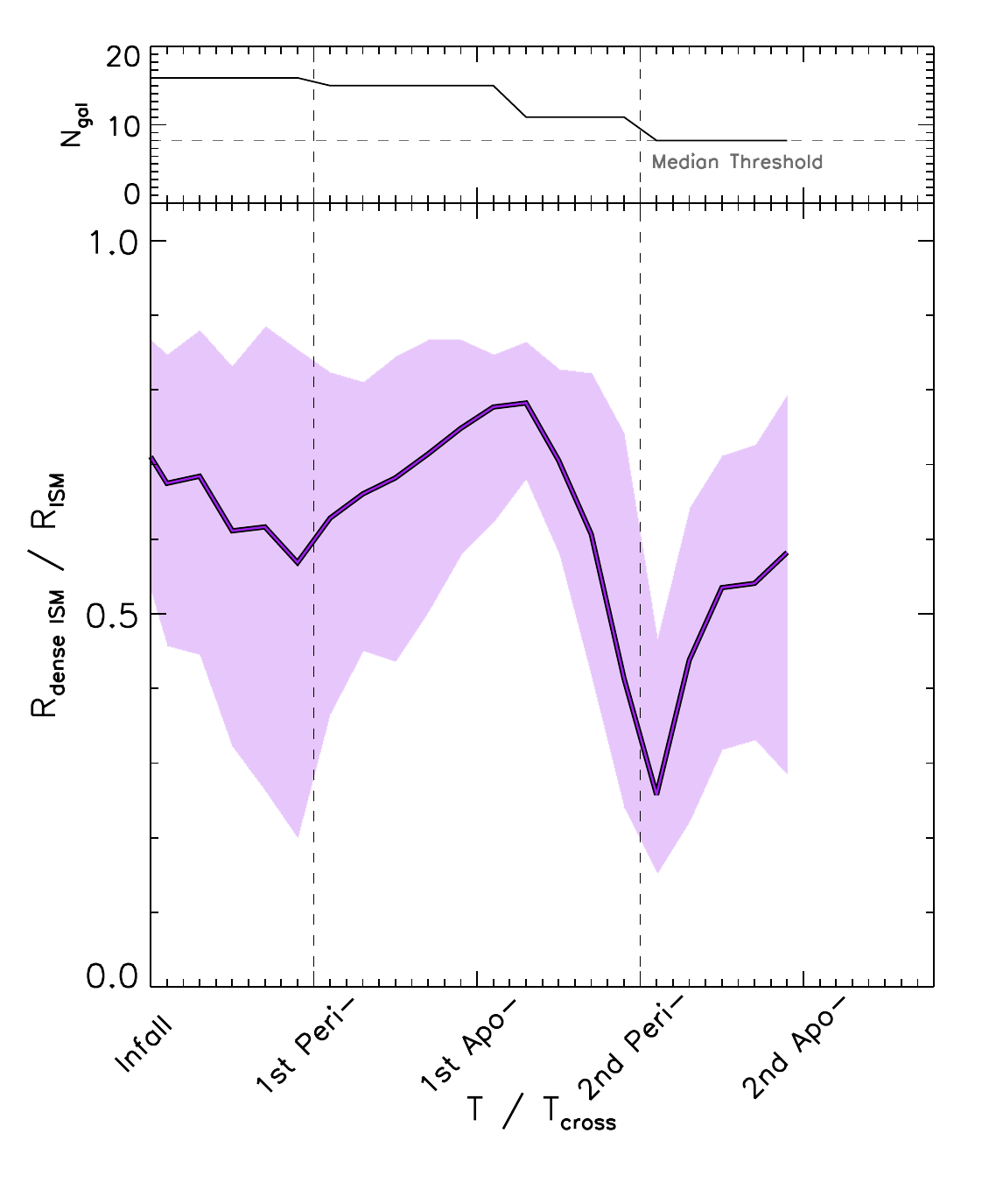}
\caption{
Fractional size evolution of massive galaxies of dense ISM gas to the total ISM gas.
The figure has the same format as in Figure \ref{fig:fig nqorbit}.
Size is defined as the half-mass radius of each gas component.
}
\label{fig:fig nqgmass2}
\end{figure}

\subsection{Origin of changes in SFE}
\label{sec:Dis-SFE}

\begin{figure*}
\centering
\includegraphics[width=0.95\textwidth]{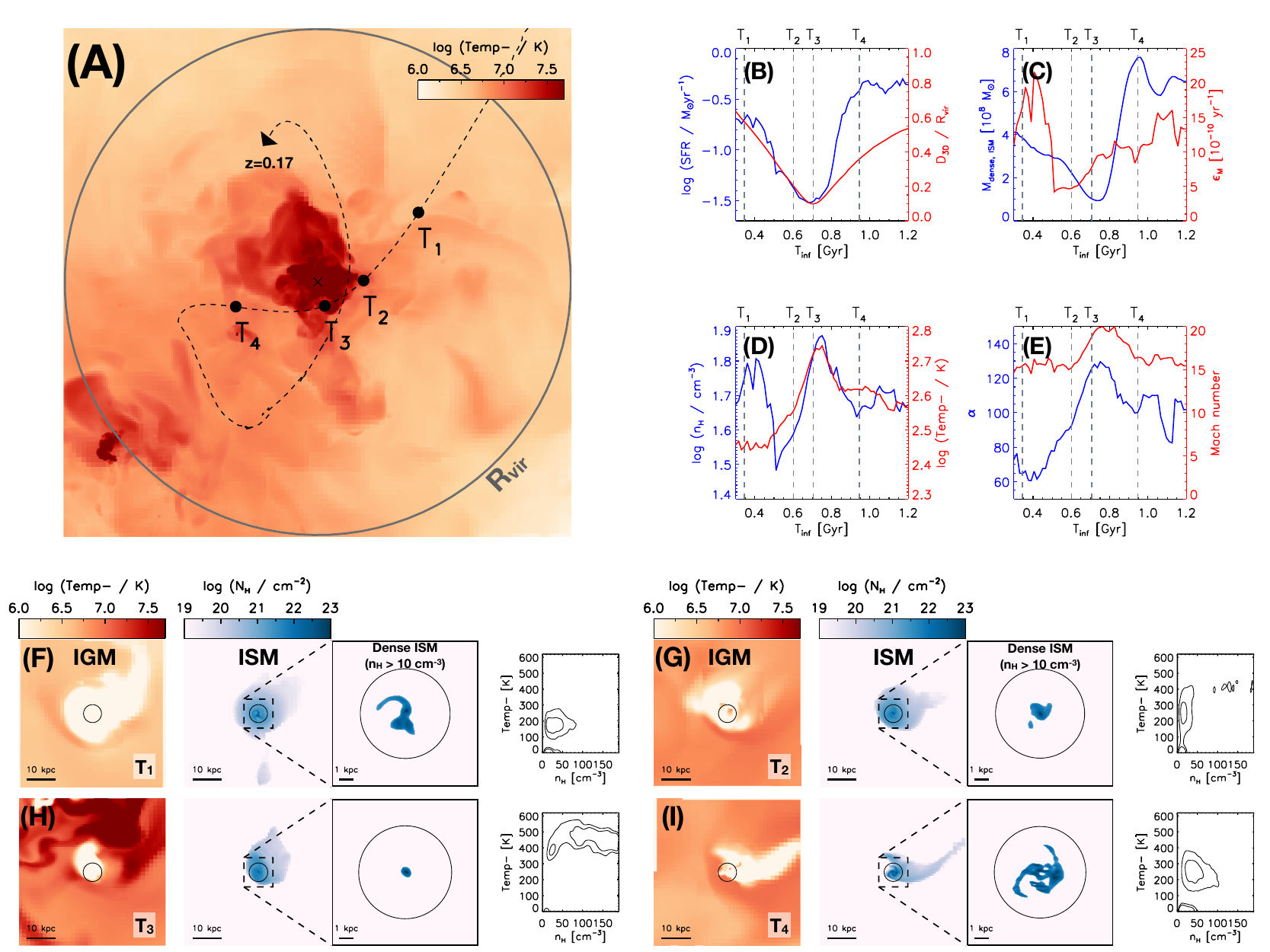}
\caption{Panel-(A): Image of the trajectory of one selected galaxy in the host group halo.
The trajectory is denoted as a black dashed line, and the background color map is the temperature distribution of the hot IGM gas when the galaxy passes through its first pericenter.
The virial radius of this group is shown with a grey solid line.
Panels (B) to (E): Evolution of galactic properties as a function of $T_{\rm inf}$ around the first pericenter passage of the selected galaxy: SFR and the normalized distance from the group center (B), the dense ISM gas mass and the mass-weighted SFE (C), the mass-weighted density and temperature of the dense ISM (D), and the virial parameter and Mach number of the dense ISM (E).
In the plots, the color-coded vertical axes correspond to the lines of the same color.
Panels (F) to (I): Temperature distribution of the surrounding IGM (left), column density of ISM and dense ISM (two middle images), and the phase diagram of ISM gas.
}
\label{fig:fig sfe}
\end{figure*}


In Section \ref{sec:Res-orbit}, the significant role of decreasing SFE in driving SF quenching is demonstrated, especially for the massive galaxy sample.
In our individual investigations of the SFE change in massive galaxies, a consistent sequence of phases becomes apparent as they orbit around their pericenter passage.
We introduce these phases by presenting one representative massive galaxy in the satellite galaxy sample with its evolution around its first pericenter passage $T_{\rm inf} = 0.3 - 1.2\,{\rm Gyr}$.

In Figure \ref{fig:fig sfe}, Panel-(A) illustrates an overview of the orbital motion of the selected massive galaxy within the host group halo.
The virial radius of the host group is shown with the grey solid line, and the orbital trajectory is depicted by the black dashed line.
This galaxy undergoes two pericenter passages, and we will focus on the four temporal epochs encompassing its first pericenter passage designated as ${\rm T}_{1}$, ${\rm T}_{2}$, ${\rm T}_{3}$, and ${\rm T}_{4}$, where ${\rm T}_{3}$ indicates the first pericenter pass.
The background color represents the temperature of the hot gas within the host group halo at ${\rm T}_{3}$.
Panel-(B) through (E) present galactic properties as a function of $T_{\rm inf}$.
Panel-(B) shows the SFR and distance to the group center, where this galaxy has a low value of SFR at the pericenter passage.
Panels-(C) through (E) show the evolution of properties of dense ISM gas that are closely involved with SF: the dense ISM gas mass and $\epsilon_{M}$ (Panel-(C)), the mass-weighted average of density and temperature of dense ISM gas cells (Panel-(D)), and the mass-weighted average of virial parameter ($\alpha$) and Mach number ($\mathcal{M}$) of dense ISM gas cells (Panel-(E)).
It is worth noting that the color-coded vertical axes in Panels-(B) to (E) correspond to the lines of the same color in the plot.
From Panel-(F) to (I), each panel includes a triad of images and an ISM phase diagram: IGM temperature map (left), ISM and dense ISM column densities (middle), and the distribution of ISM gas cells in the density-temperature plane (right).

In Panel-(A), hot IGM gas is concentrated at the group center, and this galaxy traverses this hot IGM gas during its orbital motion.
Moving to Panel-(B), this galaxy undergoes the lowest level of SF precisely at the pericenter passage (${\rm T}_{3}$), suggesting that its SF quenching may be closely related to its interactions with the hot IGM gas.
We identify three distinct phases contributing to the SF quenching.
The first phase is the ``heating'' phase, which transpires during ${\rm T}_{1}$ to ${\rm T}_{2}$.
The heating phase is characterized by a rise in temperature and a decline in the density of the dense ISM, as shown in Panel-(D).
At this phase, ISM gas temperature also increases (from $1000\,{\rm K}$ to $5000\,{\rm K}$), and thus, all gas components of this galaxy are heated.
The prominent heating source appears to be the increase of surrounding IGM gas temperature, which can be seen in the leftmost images of Panels-(F) and (G), during which adiabatic heating or induced turbulence of ISM due to IGM can be the source of such heating.
Throughout the heating phase, the dense ISM gas becomes centralized (middle images in Panels-(F) and (G)), possibly due to the fact that only the central region of this galaxy can preserve the dense ISM gas component against the external heating.
In the heating phase, both dense ISM gas mass and SFE ($\epsilon_{M}$) simultaneously decrease, thereby revealing that the SF quenching is a consequence of the decrease of the two factors.
The decrease in $\epsilon_{M}$ is mainly due to the increase of the virial parameter ($\alpha$) (Panel-(E)).
This rise in $\alpha$ implies that the dense ISM gas is becoming less virialized during the heating phase.
Therefore, in the heating phase, the temperature escalation in the surrounding IGM gas may lead to heating of the ISM, subsequently resulting in a decrease in the level of virialization, an increase in SFE, and, ultimately, the quenching of SF.

The second phase is the ``compression'' phase from ${\rm T}_{2}$ to ${\rm T}_{3}$.
In this phase, there is a significant increase in the density and temperature of the dense ISM gas (Panel-(D)).
Consequently, the overall gas component becomes denser and hotter (we refer to the contour plot in Panel-(H)).
This leads to a high level of kinetic and turbulent motion of the dense ISM as seen by the increases in the virial parameter ($\alpha$) and Mach number ($\mathcal{M}$) in Panel-(E).
While the SFE of this galaxy increases during the compression phase, in the SF model in \NH , SFE reacts differently to the increases in the virial parameter and Mach number, resulting in a highly complex pattern in other massive sample galaxies.
In addition, this phase is also characterized by the centralization and a small amount of dense ISM gas (Panels-(C) and (H)), resulting in a very low value of SFR.
Therefore, instead of heat transfer between ISM and IGM, momentum transfer appears to occur from IGM to ISM, resulting in only the central part of ISM surviving with high kinetic energy, a high Mach number, and a low value of SFR.

The last phase is the ``cooling'' phase (from ${\rm T}_{3}$ to ${\rm T}_{4}$), during which the galaxy recovers its original status.
In this phase, both the density and temperature of the dense ISM gas decrease (Panel-(D)), and the dense ISM gas mass and $\epsilon_{M}$ increase (Panel-(C)).
The dense ISM gas is mainly formed at the outskirts of the galaxy (Panel-(I)), and SF is rejuvenated from the newly formed dense ISM gas.

In summary, the SF of this galaxy is intricately regulated by interactions between the surrounding IGM and the ISM, rather than simple ISM stripping.
The interactions manifest in different ways, involving either heating the ISM (heating phase) or pushing the ISM (compression phase).
In the heated ISM, the SFE decreases as the ISM becomes hotter, and the dense ISM gas mass decreases, likely due to destruction during the heating phase.
Both factors contribute to SF quenching.
In the pushed ISM, dense ISM gas is centralized and has high density and temperature, resulting in complex variations in SFE.
After these two interactions, the influence of the surrounding IGM gas weakens, and the galaxy begins to form dense gas components at the outskirts, leading to SF rejuvenation.
We confirm that these phases are often observed in other massive galaxies, although some galaxies show only one or two phases out of the three.

\subsection{Comparison to the LG satellite galaxies}

The proximity of LG to us allows us to measure the proper motions of some of its dwarf satellites.
Despite LG having a lower virial mass than those of the \NH\ group halos, it serves as a valuable laboratory for testing our main result: orbit-related quenching of satellite galaxies.

For example, our scenario aligns well with the inferred orbital parameters and SFH of Leo I in the Milky Way halo, which has the stellar mass $\sim 5.5\times10^{6}\,M_{\odot}$ \citep[][]{McConnachie12}.
The epoch of SF quenching in Leo I, occurring about $\sim 1\,{\rm Gyr}$ ago, \citep[][]{Caputo99}, coincides with its first pericenter passage, estimated through proper motion measurements \citep[][]{Sohn13}.
In addition, with the Gaia proper motion measurements, \cite{MC20} modeled the orbital trajectories of dwarf spheroidal galaxies within the time-varying MW potential.
They compared the orbital histories of nine dwarfs with their SFH from \cite{Weisz14}, revealing that eight of them (Draco, Fornax, Leo I, Leo II, Sculptor, Ursa Minor, CVnI, and CVnII) underwent SF quenching around their first pericenter passages.
All of these dwarfs are categorized in our low-mass galaxy bin based on their stellar mass \citep[][]{McConnachie12}, sharing the same SF quenching pattern in the low-mass galaxy sample in our study.
Thus, drawing from our findings, it can be inferred that, in general, galaxies are subject to SF quenching at the first pericenter passages, while massive galaxies maintain their SF activity for extended periods.

\begin{figure}
\centering
\includegraphics[width=0.48\textwidth]{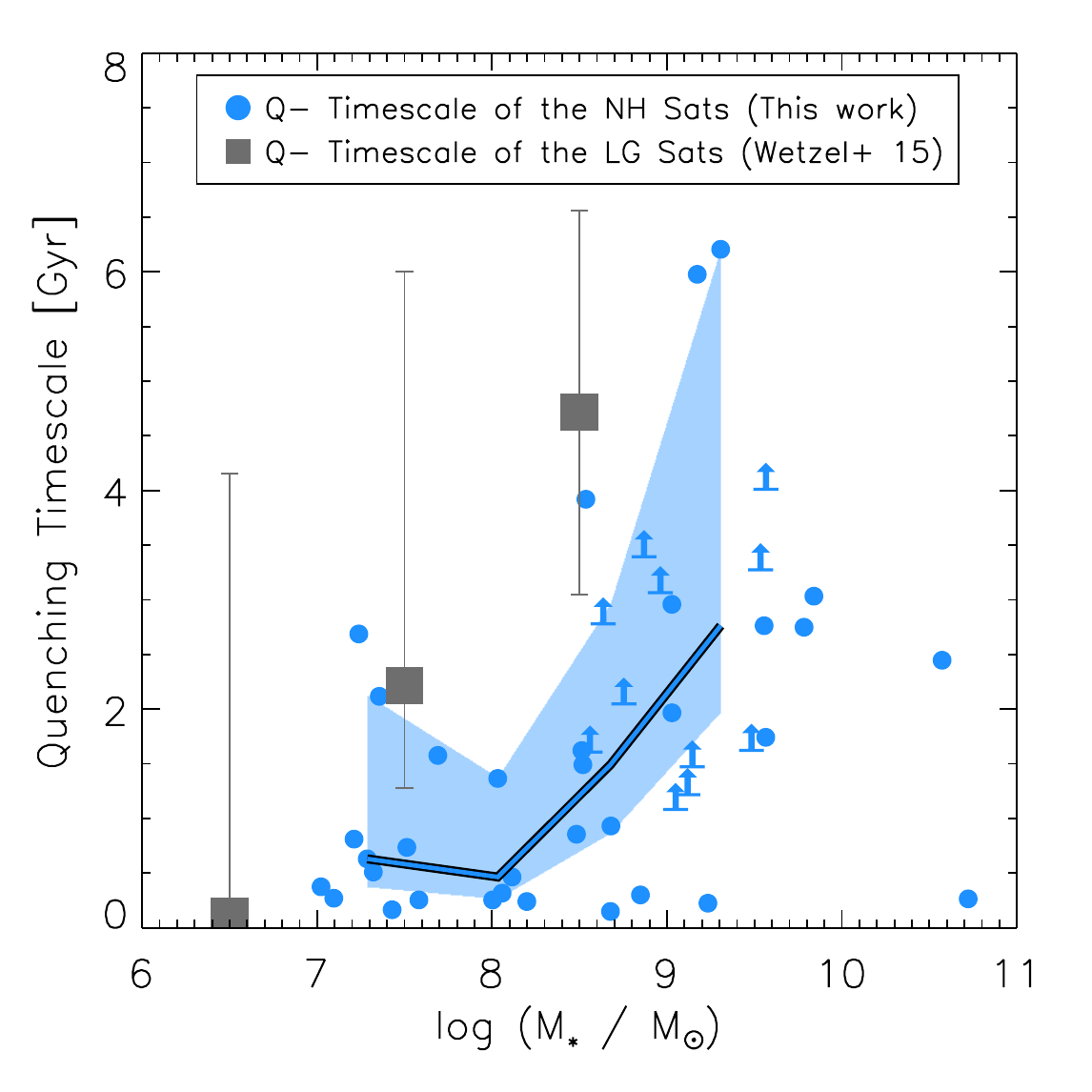}
\caption{
Quenching timescales of satellite galaxies as a function of stellar mass.
The quenching timescale is defined as the elapsed time from infall to the last quenching epoch (closed blue circles). 
Star-forming galaxies at the last snapshot only provide a lower limit for their quenching timescales, denoted by upward arrows.
The blue shaded area represents the $1^{\rm st}$ to $3^{\rm rd}$ quartile distributions of the quenching timescales, respectively, where the medians are displayed with the solid lines.
For comparison, the quenching timescales for satellite galaxies in the Local Group from \cite{Wetzel15} are also included (grey squares).
}
\label{fig:fig qtime}
\end{figure}

We also compare the quenching timescales of the satellite galaxies in \NH\ with those of the LG galaxies as shown in Figure \ref{fig:fig qtime}.
Here, the quenching timescale is defined as the elapsed time from infall to complete quenching, which is shown with closed blue circles in the figure.
The blue-shaded region indicates its $1^{\rm st}$ to $3^{\rm rd}$ quartile distribution, and the solid lines correspond to the median.
Galaxies that are not yet entirely quenched only provide a lower limit of the quenching timescales, shown with blue upward arrows.
We additionally include the quenching timescales of the LG dwarf galaxies derived in \cite{Wetzel15}, with grey squares.
Notably, LG dwarf galaxies generally have longer quenching timescales compared to the \NH\ satellites, but both share the same increasing trend with increasing stellar mass.
The relatively longer quenching timescales for the LG galaxies can be attributed to two factors.
First, the lower virial mass of the LG halo may lead to a less efficient quenching process, contributing to slower quenching.
Second, differences in quenching definitions employed between our study and \cite{Wetzel15} contributed, in which quiescent galaxies are defined as those with a low gas fraction ($M_{\rm gas}/M_{\rm star}<0.1$).
As discussed in Section \ref{sec:Res-orbit}, some of the satellite galaxies are SF quenched with a relatively high gas fraction, resulting in shorter quenching timescales in our definition compared to those in \cite{Wetzel15}.
\section{Summary and conclusions}


\begin{figure*}
\centering
\includegraphics[width=0.95\textwidth]{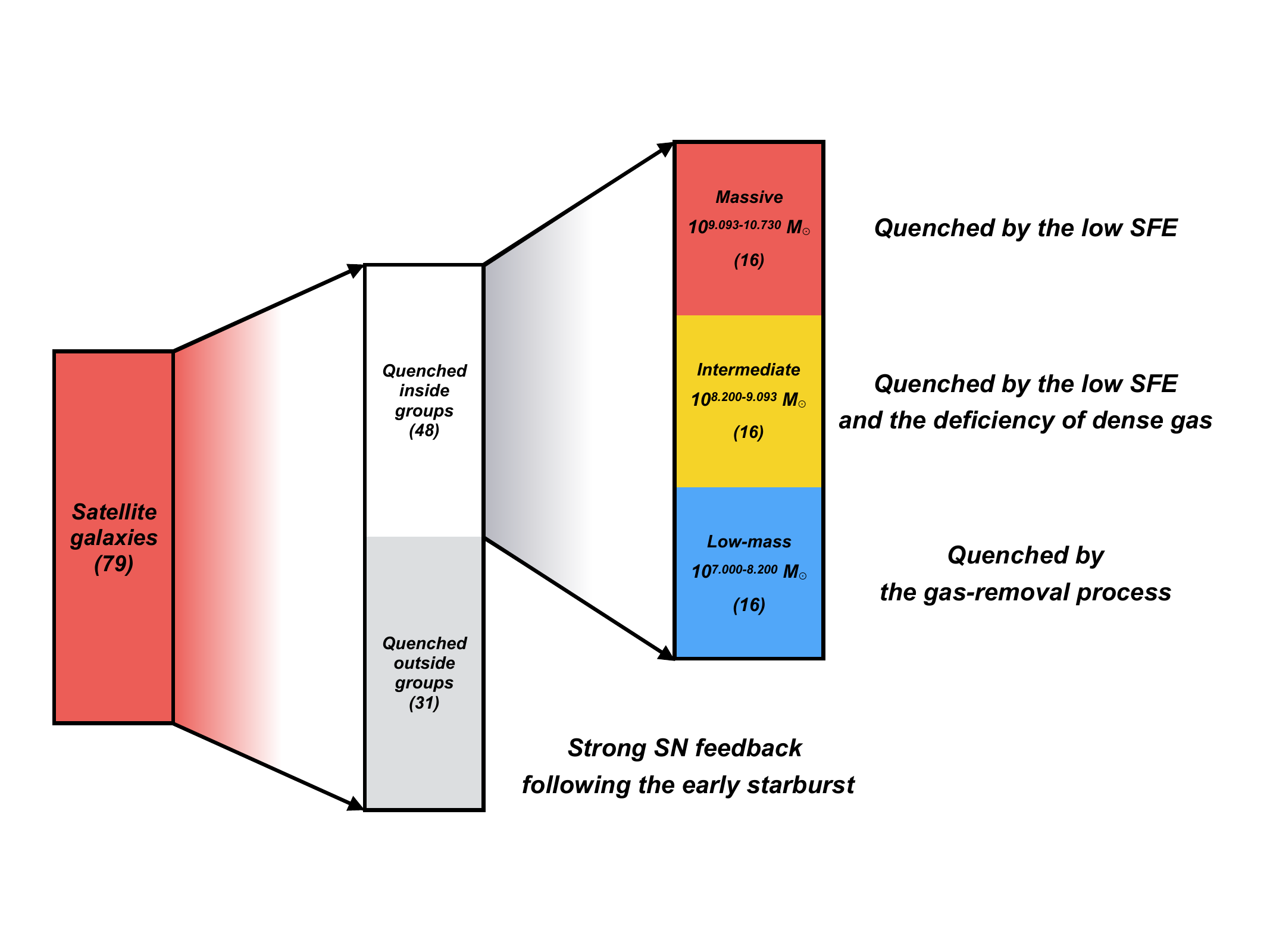}
\caption{
Summary figure about the quenching origin of satellite galaxies in the two \NH\ groups.
}
\label{fig:fig con}
\end{figure*}

In this paper, our main focus is on the origins of SF quenching of satellite galaxies ($M_{*}>10^{7} \,M_{\odot}$) in the two most massive halos in the \NH\ volume ($M_{\rm vir} = 10^{12.74}$ and $10^{12.91} \,M_{\odot}$).
Figure \ref{fig:fig con} illustrates a summary of the findings in this paper.



Among the 79 satellite galaxies used in this analysis (excluding the two BGGs), a total of 48 galaxies undergo SF quenching in their host groups, presumably due to group environmental effects.
Despite the varied SFH of these quenched galaxies, we nonetheless identify correlations between their orbital phases and the patterns in their SFH.
Specifically, galaxies have a low value of SFR at their first pericenter passage and are completely quenched at either the first or second pericenter passages.
Only massive galaxies ($>10^{9.093}\,M_{\odot}$) continue to form stars for several orbital times.

We further explore the two galactic bulk properties, the mass-weighted SFE and dense ISM gas (as a \HH\ gas tracer) mass.
We note that both are observable quantities, with high-resolution radio observations for example.
We show that the two bulk parameters in the simulation have a strong predictive power of the SFR of galaxies, as demonstrated in Figure \ref{fig:fig nqsfh}.
In our analysis presented in Figure \ref{fig:fig nqorbit}, we find that the change in SFE is the primary driver behind the SF quenching of massive galaxies.
However, for lower-mass galaxies, the role of dense ISM gas mass in determining SF activity becomes increasingly important.
In the case of low-mass galaxies ($<10^{8.2} M_{\odot}$), they become gas-deficient by interactions with the surrounding environment, leading to SF quenching.

In the context of massive galaxies at the pericenter passages, their SFE is affected mainly by interactions with surrounding hot IGM gas that envelops galaxies.
This interaction initiates a sequence of SF quenching in time.
Their ISM is initially heated as the surrounding IGM gas gets hotter.
This surge in ISM temperature results in a low SFE and the destruction of dense ISM gas.
Subsequently, the dense ISM undergoes compression and central concentration, although its overall quantity remains low during this phase.
Consequently, this compression leads to a minimum level of SF activity.
Lastly, as this ISM cools down, it transitions into a dense ISM state, rejuvenating the SF within these galaxies.
Figure \ref{fig:fig sfe} provides a visual representation of these three phases.


Changes in gas mass cannot be attributed solely to a simple ram pressure stripping scenario (Figure \ref{fig:fig nqgmass}).
Galaxies with $>10^{8.2}\,M_{\odot}$ are able to maintain their ISM gas for extended periods.
However, it is primarily the dense ISM gas that undergoes significant mass changes over orbital timescales.
Intermediate-mass galaxies undergo a gradual decline in dense ISM gas following their pericenter passages, even though they still possess a substantial amount of ISM gas at the same time.
This inability to form dense ISM gas after the first pericenter passage contributes to their subsequent slow quenching.
Low-mass galaxies, on the other hand, undergo significant gas loss after the first pericenter passage.
Remarkably, among the 16 low-mass galaxies studied, 12 of them become gas-deficient by the final snapshot ($z=0.17$).


Through our investigation, we have unveiled a physical mechanism affecting the SF activity of satellites residing in group halos.
While the definitions of cluster and group halos lack strict boundaries, we have observed a distinct quenching process in low-mass groups halos ($10^{12.7-12.9} \,M_{\odot}$) when compared to more massive cluster halos.
As a result, it becomes clear that groups have inherent differences from clusters beyond their virial masses.
This finding highlights the multifaceted nature of galaxy evolution in different environments.

In a hierarchical formation scenario, group satellite galaxies may eventually fall into larger halos, namely clusters.
Considering that group environments serve as the initial transition into dense environments for field galaxies, understanding the SF quenching in groups represents a crucial piece of the puzzle regarding the environmental effects on galaxies, namely known as pre-processing.
Hence, we provide a novel perspective on group environmental quenching and offer insights into the pre-processing phenomenon.
We acknowledge that our analysis is confined to a limited number of samples.
Therefore, our next objective is to expand the sample size and uncover additional statistical characteristics of group galaxy quenching, as well as its connection to cluster environments.

\medskip

J.R. was supported by the KASI-Yonsei Postdoctoral Fellowship and was supported by the Korea Astronomy and Space Science Institute under the R\&D program (Project No. 2023-1-830-00), supervised by the Ministry of Science and ICT.
S.K.Y. acknowledges support from the Korean National Research Foundation (2020R1A2C3003769). As the head of the group. S.K.Y. acted as the corresponding author. This work was supported by the Korean National Research Foundation (2022R1A6A1A03053472).
E.C. acknowledges support from the Korean National Research Foundation (RS-2023-00241934).
We also thank the KITP for hosting the workshop
\href{https://www.cosmicweb23.org}{`CosmicWeb23: connecting Galaxies to Cosmology at High and Low Redshift'}.
This work is partially supported by the grant \href{https://www.secular-evolution.org}{Segal} ANR-19-CE31-0017 of the French Agence Nationale de la Recherche and by the National Science Foundation under Grant No. NSF PHY-1748958.
This work was granted access to the HPC resources of CINES under the allocations c2016047637, A0020407637, and A0070402192 by Genci, KSC-2017-G2-0003 by KISTI, and as a “Grand Challenge” project granted by GENCI on the AMD Rome extension of the Joliot Curie supercomputer at TGCC.
S.P. acknowledges the support from the JSPS (Japan Society for the Promotion of Science) long-term invitation program and is grateful for the hospitality at the Department of Physics, the University of Tokyo.


\bibliography{ref}{}

\begin{thebibliography}{}
\expandafter\ifx\csname natexlab\endcsname\relax\def\natexlab#1{#1}\fi

\bibitem[{{Akins} {et~al.}(2021){Akins}, {Christensen}, {Brooks}, {Munshi},
  {Applebaum}, {Engelhardt}, \& {Chamberland}}]{Akins21}
{Akins}, H.~B., {Christensen}, C.~R., {Brooks}, A.~M., {et~al.} 2021, \apj,
  909, 139

\bibitem[{{Baes} \& {Camps}(2015)}]{BC15}
{Baes}, M., \& {Camps}, P. 2015, Astronomy and Computing, 12, 33

\bibitem[{{Bah{\'e}} \& {McCarthy}(2015)}]{BM15}
{Bah{\'e}}, Y.~M., \& {McCarthy}, I.~G. 2015, \mnras, 447, 969

\bibitem[{{Baldry} {et~al.}(2004){Baldry}, {Glazebrook}, {Brinkmann},
  {Ivezi{\'c}}, {Lupton}, {Nichol}, \& {Szalay}}]{Baldry04}
{Baldry}, I.~K., {Glazebrook}, K., {Brinkmann}, J., {et~al.} 2004, \apj, 600,
  681

\bibitem[{{Balogh} {et~al.}(2004){Balogh}, {Baldry}, {Nichol}, {Miller},
  {Bower}, \& {Glazebrook}}]{Balogh04}
{Balogh}, M.~L., {Baldry}, I.~K., {Nichol}, R., {et~al.} 2004, \apjl, 615, L101

\bibitem[{{Baxter} {et~al.}(2021){Baxter}, {Cooper}, \&
  {Fillingham}}]{Baxter21}
{Baxter}, D.~C., {Cooper}, M.~C., \& {Fillingham}, S.~P. 2021, \mnras, 503,
  1636

\bibitem[{{Bell} {et~al.}(2004){Bell}, {Wolf}, {Meisenheimer}, {Rix}, {Borch},
  {Dye}, {Kleinheinrich}, {Wisotzki}, \& {McIntosh}}]{Bell04}
{Bell}, E.~F., {Wolf}, C., {Meisenheimer}, K., {et~al.} 2004, \apj, 608, 752

\bibitem[{{Benson} {et~al.}(2003){Benson}, {Bower}, {Frenk}, {Lacey}, {Baugh},
  \& {Cole}}]{Benson03}
{Benson}, A.~J., {Bower}, R.~G., {Frenk}, C.~S., {et~al.} 2003, \apj, 599, 38

\bibitem[{{Blanton} {et~al.}(2005){Blanton}, {Eisenstein}, {Hogg}, {Schlegel},
  \& {Brinkmann}}]{Blanton05}
{Blanton}, M.~R., {Eisenstein}, D., {Hogg}, D.~W., {Schlegel}, D.~J., \&
  {Brinkmann}, J. 2005, \apj, 629, 143

\bibitem[{{Camps} \& {Baes}(2020)}]{CB20}
{Camps}, P., \& {Baes}, M. 2020, Astronomy and Computing, 31, 100381

\bibitem[{{Caputo} {et~al.}(1999){Caputo}, {Cassisi}, {Castellani}, {Marconi},
  \& {Santolamazza}}]{Caputo99}
{Caputo}, F., {Cassisi}, S., {Castellani}, M., {Marconi}, G., \&
  {Santolamazza}, P. 1999, \aj, 117, 2199

\bibitem[{{Cen} \& {Ostriker}(2006)}]{CO06}
{Cen}, R., \& {Ostriker}, J.~P. 2006, \apj, 650, 560

\bibitem[{{Chabrier}(2005)}]{Chabrier05}
{Chabrier}, G. 2005, in Astrophysics and Space Science Library, Vol. 327, The
  Initial Mass Function 50 Years Later, ed. E.~{Corbelli}, F.~{Palla}, \&
  H.~{Zinnecker}, 41

\bibitem[{{Chung} {et~al.}(2007){Chung}, {van Gorkom}, {Kenney}, \&
  {Vollmer}}]{Chung07}
{Chung}, A., {van Gorkom}, J.~H., {Kenney}, J. D.~P., \& {Vollmer}, B. 2007,
  \apjl, 659, L115

\bibitem[{{Contini} {et~al.}(2020){Contini}, {Gu}, {Ge}, {Rhee}, {Yi}, \&
  {Kang}}]{Contini20}
{Contini}, E., {Gu}, Q., {Ge}, X., {et~al.} 2020, \apj, 889, 156

\bibitem[{{Contini} {et~al.}(2019){Contini}, {Gu}, {Kang}, {Rhee}, \&
  {Yi}}]{Contini19}
{Contini}, E., {Gu}, Q., {Kang}, X., {Rhee}, J., \& {Yi}, S.~K. 2019, \apj,
  882, 167

\bibitem[{{Croton} {et~al.}(2006){Croton}, {Springel}, {White}, {De Lucia},
  {Frenk}, {Gao}, {Jenkins}, {Kauffmann}, {Navarro}, \& {Yoshida}}]{Croton06}
{Croton}, D.~J., {Springel}, V., {White}, S. D.~M., {et~al.} 2006, \mnras, 365,
  11

\bibitem[{{Dalgarno} \& {McCray}(1972)}]{DM72}
{Dalgarno}, A., \& {McCray}, R.~A. 1972, \araa, 10, 375

\bibitem[{{Darvish} {et~al.}(2016){Darvish}, {Mobasher}, {Sobral}, {Rettura},
  {Scoville}, {Faisst}, \& {Capak}}]{Darvish16}
{Darvish}, B., {Mobasher}, B., {Sobral}, D., {et~al.} 2016, \apj, 825, 113

\bibitem[{{Davies} {et~al.}(2016){Davies}, {Robotham}, {Driver}, {Alpaslan},
  {Baldry}, {Bland-Hawthorn}, {Brough}, {Brown}, {Cluver}, {Holwerda},
  {Hopkins}, {Lara-L{\'o}pez}, {Mahajan}, {Moffett}, {Owers}, \&
  {Phillipps}}]{Davies16}
{Davies}, L.~J.~M., {Robotham}, A.~S.~G., {Driver}, S.~P., {et~al.} 2016,
  \mnras, 455, 4013

\bibitem[{{De Lucia} {et~al.}(2012){De Lucia}, {Weinmann}, {Poggianti},
  {Arag{\'o}n-Salamanca}, \& {Zaritsky}}]{deLucia12}
{De Lucia}, G., {Weinmann}, S., {Poggianti}, B.~M., {Arag{\'o}n-Salamanca}, A.,
  \& {Zaritsky}, D. 2012, \mnras, 423, 1277

\bibitem[{{Dekel} \& {Silk}(1986)}]{DS86}
{Dekel}, A., \& {Silk}, J. 1986, \apj, 303, 39

\bibitem[{{Dubois} {et~al.}(2014){Dubois}, {Pichon}, {Welker}, {Le Borgne},
  {Devriendt}, {Laigle}, {Codis}, {Pogosyan}, {Arnouts}, {Benabed}, {Bertin},
  {Blaizot}, {Bouchet}, {Cardoso}, {Colombi}, {de Lapparent}, {Desjacques},
  {Gavazzi}, {Kassin}, {Kimm}, {McCracken}, {Milliard}, {Peirani}, {Prunet},
  {Rouberol}, {Silk}, {Slyz}, {Sousbie}, {Teyssier}, {Tresse}, {Treyer},
  {Vibert}, \& {Volonteri}}]{Dubois14}
{Dubois}, Y., {Pichon}, C., {Welker}, C., {et~al.} 2014, \mnras, 444, 1453

\bibitem[{{Dubois} {et~al.}(2021){Dubois}, {Beckmann}, {Bournaud}, {Choi},
  {Devriendt}, {Jackson}, {Kaviraj}, {Kimm}, {Kraljic}, {Laigle}, {Martin},
  {Park}, {Peirani}, {Pichon}, {Volonteri}, \& {Yi}}]{Dubois21}
{Dubois}, Y., {Beckmann}, R., {Bournaud}, F., {et~al.} 2021, \aap, 651, A109

\bibitem[{{Einasto} {et~al.}(1974){Einasto}, {Saar}, {Kaasik}, \&
  {Chernin}}]{Einasto74}
{Einasto}, J., {Saar}, E., {Kaasik}, A., \& {Chernin}, A.~D. 1974, \nat, 252,
  111

\bibitem[{{Elahi} {et~al.}(2019{\natexlab{a}}){Elahi}, {Ca{\~n}as}, {Poulton},
  {Tobar}, {Willis}, {Lagos}, {Power}, \& {Robotham}}]{Elahi19a}
{Elahi}, P.~J., {Ca{\~n}as}, R., {Poulton}, R. J.~J., {et~al.}
  2019{\natexlab{a}}, \pasa, 36, e021

\bibitem[{{Elahi} {et~al.}(2019{\natexlab{b}}){Elahi}, {Poulton}, {Tobar},
  {Ca{\~n}as}, {Lagos}, {Power}, \& {Robotham}}]{Elahi19b}
{Elahi}, P.~J., {Poulton}, R. J.~J., {Tobar}, R.~J., {et~al.}
  2019{\natexlab{b}}, \pasa, 36, e028

\bibitem[{{Faber} {et~al.}(2007){Faber}, {Willmer}, {Wolf}, {Koo}, {Weiner},
  {Newman}, {Im}, {Coil}, {Conroy}, {Cooper}, {Davis}, {Finkbeiner}, {Gerke},
  {Gebhardt}, {Groth}, {Guhathakurta}, {Harker}, {Kaiser}, {Kassin},
  {Kleinheinrich}, {Konidaris}, {Kron}, {Lin}, {Luppino}, {Madgwick},
  {Meisenheimer}, {Noeske}, {Phillips}, {Sarajedini}, {Schiavon}, {Simard},
  {Szalay}, {Vogt}, \& {Yan}}]{Faber07}
{Faber}, S.~M., {Willmer}, C.~N.~A., {Wolf}, C., {et~al.} 2007, \apj, 665, 265

\bibitem[{{Federrath} \& {Klessen}(2012)}]{FK12}
{Federrath}, C., \& {Klessen}, R.~S. 2012, \apj, 761, 156

\bibitem[{{Fillingham} {et~al.}(2016){Fillingham}, {Cooper}, {Pace},
  {Boylan-Kolchin}, {Bullock}, {Garrison-Kimmel}, \& {Wheeler}}]{Fillingham16}
{Fillingham}, S.~P., {Cooper}, M.~C., {Pace}, A.~B., {et~al.} 2016, \mnras,
  463, 1916

\bibitem[{{Fillingham} {et~al.}(2015){Fillingham}, {Cooper}, {Wheeler},
  {Garrison-Kimmel}, {Boylan-Kolchin}, \& {Bullock}}]{Fillingham15}
{Fillingham}, S.~P., {Cooper}, M.~C., {Wheeler}, C., {et~al.} 2015, \mnras,
  454, 2039

\bibitem[{{Foltz} {et~al.}(2018){Foltz}, {Wilson}, {Muzzin}, {Cooper},
  {Nantais}, {van der Burg}, {Cerulo}, {Chan}, {Fillingham}, {Surace}, {Webb},
  {Noble}, {Lacy}, {McDonald}, {Rudnick}, {Lidman}, {Demarco},
  {Hlavacek-Larrondo}, {Yee}, {Perlmutter}, \& {Hayden}}]{Foltz18}
{Foltz}, R., {Wilson}, G., {Muzzin}, A., {et~al.} 2018, \apj, 866, 136

\bibitem[{{Font} {et~al.}(2022){Font}, {McCarthy}, {Belokurov}, {Brown}, \&
  {Stafford}}]{Font22}
{Font}, A.~S., {McCarthy}, I.~G., {Belokurov}, V., {Brown}, S.~T., \&
  {Stafford}, S.~G. 2022, \mnras, 511, 1544

\bibitem[{{Franx} {et~al.}(2008){Franx}, {van Dokkum}, {F{\"o}rster Schreiber},
  {Wuyts}, {Labb{\'e}}, \& {Toft}}]{Franx08}
{Franx}, M., {van Dokkum}, P.~G., {F{\"o}rster Schreiber}, N.~M., {et~al.}
  2008, \apj, 688, 770

\bibitem[{{Geha} {et~al.}(2012){Geha}, {Blanton}, {Yan}, \& {Tinker}}]{Geha12}
{Geha}, M., {Blanton}, M.~R., {Yan}, R., \& {Tinker}, J.~L. 2012, \apj, 757, 85

\bibitem[{{Grcevich} \& {Putman}(2009)}]{GP09}
{Grcevich}, J., \& {Putman}, M.~E. 2009, \apj, 696, 385

\bibitem[{{Gunn} \& {Gott}(1972)}]{GG72}
{Gunn}, J.~E., \& {Gott}, J.~Richard, I. 1972, \apj, 176, 1

\bibitem[{{Haardt} \& {Madau}(1996)}]{HM96}
{Haardt}, F., \& {Madau}, P. 1996, \apj, 461, 20

\bibitem[{{Haines} {et~al.}(2008){Haines}, {Gargiulo}, \&
  {Merluzzi}}]{Haines08}
{Haines}, C.~P., {Gargiulo}, A., \& {Merluzzi}, P. 2008, \mnras, 385, 1201

\bibitem[{{Han} {et~al.}(2018){Han}, {Smith}, {Choi}, {Cortese}, {Catinella},
  {Contini}, \& {Yi}}]{Han18}
{Han}, S., {Smith}, R., {Choi}, H., {et~al.} 2018, \apj, 866, 78

\bibitem[{{Hogg} {et~al.}(2003){Hogg}, {Blanton}, {Eisenstein}, {Gunn},
  {Schlegel}, {Zehavi}, {Bahcall}, {Brinkmann}, {Csabai}, {Schneider},
  {Weinberg}, \& {York}}]{Hogg03}
{Hogg}, D.~W., {Blanton}, M.~R., {Eisenstein}, D.~J., {et~al.} 2003, \apjl,
  585, L5

\bibitem[{{Jackson} {et~al.}(2021{\natexlab{a}}){Jackson}, {Kaviraj}, {Martin},
  {Devriendt}, {Slyz}, {Silk}, {Dubois}, {Yi}, {Pichon}, {Volonteri}, {Choi},
  {Kimm}, {Kraljic}, \& {Peirani}}]{Jackson21a}
{Jackson}, R.~A., {Kaviraj}, S., {Martin}, G., {et~al.} 2021{\natexlab{a}},
  \mnras, 502, 1785

\bibitem[{{Jackson} {et~al.}(2021{\natexlab{b}}){Jackson}, {Martin}, {Kaviraj},
  {Rams{\o}y}, {Devriendt}, {Sedgwick}, {Laigle}, {Choi}, {Beckmann},
  {Volonteri}, {Dubois}, {Pichon}, {Yi}, {Slyz}, {Kraljic}, {Kimm}, {Peirani},
  \& {Baldry}}]{Jackson21b}
{Jackson}, R.~A., {Martin}, G., {Kaviraj}, S., {et~al.} 2021{\natexlab{b}},
  \mnras, 502, 4262

\bibitem[{{Jang} {et~al.}(2023){Jang}, {Yi}, {Dubois}, {Rhee}, {Pichon},
  {Kimm}, {Devriendt}, {Volonteri}, {Kaviraj}, {Peirani}, {Oh}, \&
  {Croom}}]{Jang23}
{Jang}, J.~K., {Yi}, S.~K., {Dubois}, Y., {et~al.} 2023, \apj, 950, 4

\bibitem[{{Jeon} {et~al.}(2022){Jeon}, {Yi}, {Dubois}, {Chung}, {Devriendt},
  {Han}, {Jackson}, {Kimm}, {Pichon}, \& {Rhee}}]{Jeon22}
{Jeon}, S., {Yi}, S.~K., {Dubois}, Y., {et~al.} 2022, \apj, 941, 5

\bibitem[{{Karachentsev} \& {Kaisina}(2013)}]{Karachentsev13}
{Karachentsev}, I.~D., \& {Kaisina}, E.~I. 2013, \aj, 146, 46

\bibitem[{{Karunakaran} {et~al.}(2021){Karunakaran}, {Spekkens}, {Oman},
  {Simpson}, {Fattahi}, {Sand}, {Bennet}, {Crnojevi{\'c}}, {Frenk},
  {G{\'o}mez}, {Grand}, {Jones}, {Marinacci}, {Mutlu-Pakdil}, {Navarro}, \&
  {Zaritsky}}]{Karunakaran21}
{Karunakaran}, A., {Spekkens}, K., {Oman}, K.~A., {et~al.} 2021, \apjl, 916,
  L19

\bibitem[{{Kauffmann} {et~al.}(2004){Kauffmann}, {White}, {Heckman},
  {M{\'e}nard}, {Brinchmann}, {Charlot}, {Tremonti}, \&
  {Brinkmann}}]{Kauffmann04}
{Kauffmann}, G., {White}, S. D.~M., {Heckman}, T.~M., {et~al.} 2004, \mnras,
  353, 713

\bibitem[{{Kimm} \& {Cen}(2014)}]{KC14}
{Kimm}, T., \& {Cen}, R. 2014, \apj, 788, 121

\bibitem[{{Kimm} {et~al.}(2015){Kimm}, {Cen}, {Devriendt}, {Dubois}, \&
  {Slyz}}]{Kimm15}
{Kimm}, T., {Cen}, R., {Devriendt}, J., {Dubois}, Y., \& {Slyz}, A. 2015,
  \mnras, 451, 2900

\bibitem[{{Kimm} {et~al.}(2017){Kimm}, {Katz}, {Haehnelt}, {Rosdahl},
  {Devriendt}, \& {Slyz}}]{Kimm17}
{Kimm}, T., {Katz}, H., {Haehnelt}, M., {et~al.} 2017, \mnras, 466, 4826

\bibitem[{{Kimm} {et~al.}(2009){Kimm}, {Somerville}, {Yi}, {van den Bosch},
  {Salim}, {Fontanot}, {Monaco}, {Mo}, {Pasquali}, {Rich}, \& {Yang}}]{Kimm09}
{Kimm}, T., {Somerville}, R.~S., {Yi}, S.~K., {et~al.} 2009, \mnras, 394, 1131

\bibitem[{{Komatsu} {et~al.}(2011){Komatsu}, {Smith}, {Dunkley}, {Bennett},
  {Gold}, {Hinshaw}, {Jarosik}, {Larson}, {Nolta}, {Page}, {Spergel},
  {Halpern}, {Hill}, {Kogut}, {Limon}, {Meyer}, {Odegard}, {Tucker}, {Weiland},
  {Wollack}, \& {Wright}}]{Komatsu11}
{Komatsu}, E., {Smith}, K.~M., {Dunkley}, J., {et~al.} 2011, \apjs, 192, 18

\bibitem[{{Kraljic} {et~al.}(2024){Kraljic}, {Renaud}, {Dubois}, {Pichon},
  {Agertz}, {Andersson}, {Devriendt}, {Freundlich}, {Kaviraj}, {Kimm},
  {Martin}, {Peirani}, {Segovia Otero}, {Volonteri}, \& {Yi}}]{Katarina24}
{Kraljic}, K., {Renaud}, F., {Dubois}, Y., {et~al.} 2024, \aap, 682, A50

\bibitem[{{Larson}(1974)}]{Larson74}
{Larson}, R.~B. 1974, \mnras, 169, 229

\bibitem[{{Lee} \& {Chung}(2018)}]{LC18}
{Lee}, B., \& {Chung}, A. 2018, \apjl, 866, L10

\bibitem[{{Lee} {et~al.}(2022){Lee}, {Wang}, {Chung}, {Ho}, {Wang},
  {Michiyama}, {Molina}, {Kim}, {Shao}, {Kilborn}, {Wang}, {Lin}, {Kim},
  {Catinella}, {Cortese}, {Deg}, {Denes}, {Elagali}, {For}, {Kleiner},
  {Koribalski}, {Lee-Waddell}, {Rhee}, {Spekkens}, {Westmeier}, {Wong},
  {Bigiel}, {Bosma}, {Holwerda}, {van der Hulst}, {Roychowdhury},
  {Verdes-Montenegro}, \& {Zwaan}}]{Lee22}
{Lee}, B., {Wang}, J., {Chung}, A., {et~al.} 2022, \apjs, 262, 31

\bibitem[{{Limousin} {et~al.}(2009){Limousin}, {Sommer-Larsen}, {Natarajan}, \&
  {Milvang-Jensen}}]{Limousin09}
{Limousin}, M., {Sommer-Larsen}, J., {Natarajan}, P., \& {Milvang-Jensen}, B.
  2009, \apj, 696, 1771

\bibitem[{{Lotz} {et~al.}(2019){Lotz}, {Remus}, {Dolag}, {Biviano}, \&
  {Burkert}}]{Lotz19}
{Lotz}, M., {Remus}, R.-S., {Dolag}, K., {Biviano}, A., \& {Burkert}, A. 2019,
  \mnras, 488, 5370

\bibitem[{{Mao} {et~al.}(2021){Mao}, {Geha}, {Wechsler}, {Weiner}, {Tollerud},
  {Nadler}, \& {Kallivayalil}}]{Mao21}
{Mao}, Y.-Y., {Geha}, M., {Wechsler}, R.~H., {et~al.} 2021, \apj, 907, 85

\bibitem[{{Martin} {et~al.}(2007){Martin}, {Wyder}, {Schiminovich}, {Barlow},
  {Forster}, {Friedman}, {Morrissey}, {Neff}, {Seibert}, {Small}, {Welsh},
  {Bianchi}, {Donas}, {Heckman}, {Lee}, {Madore}, {Milliard}, {Rich}, {Szalay},
  \& {Yi}}]{Martin07}
{Martin}, D.~C., {Wyder}, T.~K., {Schiminovich}, D., {et~al.} 2007, \apjs, 173,
  342

\bibitem[{{Martin} {et~al.}(2022){Martin}, {Bazkiaei}, {Spavone}, {Iodice},
  {Mihos}, {Montes}, {Benavides}, {Brough}, {Carlin}, {Collins}, {Duc},
  {G{\'o}mez}, {Galaz}, {Hern{\'a}ndez-Toledo}, {Jackson}, {Kaviraj}, {Knapen},
  {Mart{\'\i}nez-Lombilla}, {McGee}, {O'Ryan}, {Prole}, {Rich}, {Rom{\'a}n},
  {Shah}, {Starkenburg}, {Watkins}, {Zaritsky}, {Pichon}, {Armus}, {Bianconi},
  {Buitrago}, {Bus{\'a}}, {Davis}, {Demarco}, {Desmons}, {Garc{\'\i}a},
  {Graham}, {Holwerda}, {Hon}, {Khalid}, {Klehammer}, {Klutse}, {Lazar},
  {Nair}, {Noakes-Kettel}, {Rutkowski}, {Saha}, {Sahu}, {Sola},
  {V{\'a}zquez-Mata}, {Vera-Casanova}, \& {Yoon}}]{Martin22}
{Martin}, G., {Bazkiaei}, A.~E., {Spavone}, M., {et~al.} 2022, \mnras, 513,
  1459

\bibitem[{{Mateo}(1998)}]{Mateo98}
{Mateo}, M.~L. 1998, \araa, 36, 435

\bibitem[{{McConnachie}(2012)}]{McConnachie12}
{McConnachie}, A.~W. 2012, \aj, 144, 4

\bibitem[{{Mei} {et~al.}(2009){Mei}, {Holden}, {Blakeslee}, {Ford}, {Franx},
  {Homeier}, {Illingworth}, {Jee}, {Overzier}, {Postman}, {Rosati}, {Van der
  Wel}, \& {Bartlett}}]{Mei09}
{Mei}, S., {Holden}, B.~P., {Blakeslee}, J.~P., {et~al.} 2009, \apj, 690, 42

\bibitem[{{Mihos}(2004)}]{Mihos04}
{Mihos}, J.~C. 2004, in Clusters of Galaxies: Probes of Cosmological Structure
  and Galaxy Evolution, ed. J.~S. {Mulchaey}, A.~{Dressler}, \& A.~{Oemler},
  277

\bibitem[{{Miyoshi} \& {Chiba}(2020)}]{MC20}
{Miyoshi}, T., \& {Chiba}, M. 2020, \apj, 905, 109

\bibitem[{{Mok} {et~al.}(2014){Mok}, {Balogh}, {McGee}, {Wilman}, {Finoguenov},
  {Tanaka}, {Bower}, {Hou}, {Mulchaey}, \& {Parker}}]{Mock14}
{Mok}, A., {Balogh}, M.~L., {McGee}, S.~L., {et~al.} 2014, \mnras, 438, 3070

\bibitem[{{Moore} {et~al.}(1996){Moore}, {Katz}, {Lake}, {Dressler}, \&
  {Oemler}}]{Moore96}
{Moore}, B., {Katz}, N., {Lake}, G., {Dressler}, A., \& {Oemler}, A. 1996,
  \nat, 379, 613

\bibitem[{{Oman} {et~al.}(2021){Oman}, {Bah{\'e}}, {Healy}, {Hess}, {Hudson},
  \& {Verheijen}}]{Oman21}
{Oman}, K.~A., {Bah{\'e}}, Y.~M., {Healy}, J., {et~al.} 2021, \mnras, 501, 5073

\bibitem[{{Park} {et~al.}(2022){Park}, {Tacchella}, {Nelson}, {Hernquist},
  {Weinberger}, {Diemer}, {Nelson}, {Pillepich}, {Marinacci}, \&
  {Vogelsberger}}]{Park22}
{Park}, M., {Tacchella}, S., {Nelson}, E.~J., {et~al.} 2022, \mnras, 515, 213

\bibitem[{{Peng} {et~al.}(2010){Peng}, {Lilly}, {Kova{\v{c}}}, {Bolzonella},
  {Pozzetti}, {Renzini}, {Zamorani}, {Ilbert}, {Knobel}, {Iovino}, {Maier},
  {Cucciati}, {Tasca}, {Carollo}, {Silverman}, {Kampczyk}, {de Ravel},
  {Sanders}, {Scoville}, {Contini}, {Mainieri}, {Scodeggio}, {Kneib}, {Le
  F{\`e}vre}, {Bardelli}, {Bongiorno}, {Caputi}, {Coppa}, {de la Torre},
  {Franzetti}, {Garilli}, {Lamareille}, {Le Borgne}, {Le Brun}, {Mignoli},
  {Perez Montero}, {Pello}, {Ricciardelli}, {Tanaka}, {Tresse}, {Vergani},
  {Welikala}, {Zucca}, {Oesch}, {Abbas}, {Barnes}, {Bordoloi}, {Bottini},
  {Cappi}, {Cassata}, {Cimatti}, {Fumana}, {Hasinger}, {Koekemoer},
  {Leauthaud}, {Maccagni}, {Marinoni}, {McCracken}, {Memeo}, {Meneux}, {Nair},
  {Porciani}, {Presotto}, \& {Scaramella}}]{Peng10}
{Peng}, Y.-j., {Lilly}, S.~J., {Kova{\v{c}}}, K., {et~al.} 2010, \apj, 721, 193

\bibitem[{{Putman} {et~al.}(2021){Putman}, {Zheng}, {Price-Whelan}, {Grcevich},
  {Johnson}, {Tollerud}, \& {Peek}}]{Putman21}
{Putman}, M.~E., {Zheng}, Y., {Price-Whelan}, A.~M., {et~al.} 2021, \apj, 913,
  53

\bibitem[{{Reddish} {et~al.}(2022){Reddish}, {Kraljic}, {Petersen}, {Tep},
  {Dubois}, {Pichon}, {Peirani}, {Bournaud}, {Choi}, {Devriendt}, {Jackson},
  {Martin}, {Park}, {Volonteri}, \& {Yi}}]{Reddish22}
{Reddish}, J., {Kraljic}, K., {Petersen}, M.~S., {et~al.} 2022, \mnras, 512,
  160

\bibitem[{{Rhee} {et~al.}(2022){Rhee}, {Elahi}, \& {Yi}}]{Rhee22}
{Rhee}, J., {Elahi}, P., \& {Yi}, S.~K. 2022, \apj, 927, 129

\bibitem[{{Rhee} {et~al.}(2020){Rhee}, {Smith}, {Choi}, {Contini}, {Jung},
  {Han}, \& {Yi}}]{Rhee20}
{Rhee}, J., {Smith}, R., {Choi}, H., {et~al.} 2020, \apjs, 247, 45

\bibitem[{{Rhee} {et~al.}(2017){Rhee}, {Smith}, {Choi}, {Yi}, {Jaff{\'e}},
  {Candlish}, \& {S{\'a}nchez-J{\'a}nssen}}]{Rhee17}
---. 2017, \apj, 843, 128

\bibitem[{{Salim} {et~al.}(2007){Salim}, {Rich}, {Charlot}, {Brinchmann},
  {Johnson}, {Schiminovich}, {Seibert}, {Mallery}, {Heckman}, {Forster},
  {Friedman}, {Martin}, {Morrissey}, {Neff}, {Small}, {Wyder}, {Bianchi},
  {Donas}, {Lee}, {Madore}, {Milliard}, {Szalay}, {Welsh}, \& {Yi}}]{Salim07}
{Salim}, S., {Rich}, R.~M., {Charlot}, S., {et~al.} 2007, \apjs, 173, 267

\bibitem[{{Samuel} {et~al.}(2022){Samuel}, {Wetzel}, {Santistevan}, {Tollerud},
  {Moreno}, {Boylan-Kolchin}, {Bailin}, \& {Pardasani}}]{Samuel22}
{Samuel}, J., {Wetzel}, A., {Santistevan}, I., {et~al.} 2022, \mnras, 514, 5276

\bibitem[{{Schawinski} {et~al.}(2014){Schawinski}, {Urry}, {Simmons},
  {Fortson}, {Kaviraj}, {Keel}, {Lintott}, {Masters}, {Nichol}, {Sarzi},
  {Skibba}, {Treister}, {Willett}, {Wong}, \& {Yi}}]{Schawinski14}
{Schawinski}, K., {Urry}, C.~M., {Simmons}, B.~D., {et~al.} 2014, \mnras, 440,
  889

\bibitem[{{Schaye} {et~al.}(2015){Schaye}, {Crain}, {Bower}, {Furlong},
  {Schaller}, {Theuns}, {Dalla Vecchia}, {Frenk}, {McCarthy}, {Helly},
  {Jenkins}, {Rosas-Guevara}, {White}, {Baes}, {Booth}, {Camps}, {Navarro},
  {Qu}, {Rahmati}, {Sawala}, {Thomas}, \& {Trayford}}]{Schaye15}
{Schaye}, J., {Crain}, R.~A., {Bower}, R.~G., {et~al.} 2015, \mnras, 446, 521

\bibitem[{{Silverman} {et~al.}(2008){Silverman}, {Mainieri}, {Lehmer},
  {Alexander}, {Bauer}, {Bergeron}, {Brandt}, {Gilli}, {Hasinger}, {Schneider},
  {Tozzi}, {Vignali}, {Koekemoer}, {Miyaji}, {Popesso}, {Rosati}, \&
  {Szokoly}}]{Silverman08}
{Silverman}, J.~D., {Mainieri}, V., {Lehmer}, B.~D., {et~al.} 2008, \apj, 675,
  1025

\bibitem[{{Simpson} {et~al.}(2018){Simpson}, {Grand}, {G{\'o}mez}, {Marinacci},
  {Pakmor}, {Springel}, {Campbell}, \& {Frenk}}]{Simpson18}
{Simpson}, C.~M., {Grand}, R. J.~J., {G{\'o}mez}, F.~A., {et~al.} 2018, \mnras,
  478, 548

\bibitem[{{Smith} {et~al.}(2016){Smith}, {Choi}, {Lee}, {Rhee},
  {Sanchez-Janssen}, \& {Yi}}]{Smith16}
{Smith}, R., {Choi}, H., {Lee}, J., {et~al.} 2016, \apj, 833, 109

\bibitem[{{Sohn} {et~al.}(2013){Sohn}, {Besla}, {van der Marel},
  {Boylan-Kolchin}, {Majewski}, \& {Bullock}}]{Sohn13}
{Sohn}, S.~T., {Besla}, G., {van der Marel}, R.~P., {et~al.} 2013, \apj, 768,
  139

\bibitem[{{Strateva} {et~al.}(2001){Strateva}, {Ivezi{\'c}}, {Knapp},
  {Narayanan}, {Strauss}, {Gunn}, {Lupton}, {Schlegel}, {Bahcall}, {Brinkmann},
  {Brunner}, {Budav{\'a}ri}, {Csabai}, {Castander}, {Doi}, {Fukugita},
  {Gy{\H{o}}ry}, {Hamabe}, {Hennessy}, {Ichikawa}, {Kunszt}, {Lamb}, {McKay},
  {Okamura}, {Racusin}, {Sekiguchi}, {Schneider}, {Shimasaku}, \&
  {York}}]{Strateva01}
{Strateva}, I., {Ivezi{\'c}}, {\v{Z}}., {Knapp}, G.~R., {et~al.} 2001, \aj,
  122, 1861

\bibitem[{{Sutherland} \& {Dopita}(1993)}]{SD93}
{Sutherland}, R.~S., \& {Dopita}, M.~A. 1993, \apjs, 88, 253

\bibitem[{{Teyssier}(2002)}]{Teyssier02}
{Teyssier}, R. 2002, \aap, 385, 337

\bibitem[{{Tolstoy} {et~al.}(2009){Tolstoy}, {Hill}, \& {Tosi}}]{Tolstoy09}
{Tolstoy}, E., {Hill}, V., \& {Tosi}, M. 2009, \araa, 47, 371

\bibitem[{{Torrey} {et~al.}(2012){Torrey}, {Vogelsberger}, {Sijacki},
  {Springel}, \& {Hernquist}}]{Torrey12}
{Torrey}, P., {Vogelsberger}, M., {Sijacki}, D., {Springel}, V., \&
  {Hernquist}, L. 2012, \mnras, 427, 2224

\bibitem[{{Weisz} {et~al.}(2014){Weisz}, {Dolphin}, {Skillman}, {Holtzman},
  {Gilbert}, {Dalcanton}, \& {Williams}}]{Weisz14}
{Weisz}, D.~R., {Dolphin}, A.~E., {Skillman}, E.~D., {et~al.} 2014, \apj, 789,
  147

\bibitem[{{Weisz} {et~al.}(2015){Weisz}, {Dolphin}, {Skillman}, {Holtzman},
  {Gilbert}, {Dalcanton}, \& {Williams}}]{Weisz15}
---. 2015, \apj, 804, 136

\bibitem[{{Wetzel} {et~al.}(2016){Wetzel}, {Hopkins}, {Kim},
  {Faucher-Gigu{\`e}re}, {Kere{\v{s}}}, \& {Quataert}}]{Wetzel16}
{Wetzel}, A.~R., {Hopkins}, P.~F., {Kim}, J.-h., {et~al.} 2016, \apjl, 827, L23

\bibitem[{{Wetzel} {et~al.}(2013){Wetzel}, {Tinker}, {Conroy}, \& {van den
  Bosch}}]{Wetzel13}
{Wetzel}, A.~R., {Tinker}, J.~L., {Conroy}, C., \& {van den Bosch}, F.~C. 2013,
  \mnras, 432, 336

\bibitem[{{Wetzel} {et~al.}(2015){Wetzel}, {Tollerud}, \& {Weisz}}]{Wetzel15}
{Wetzel}, A.~R., {Tollerud}, E.~J., \& {Weisz}, D.~R. 2015, \apjl, 808, L27

\bibitem[{{Yi} {et~al.}(2023){Yi}, {Jang}, {Devriendt}, {Dubois}, {Han},
  {Kimm}, {Kraljic}, {Park}, {Peirani}, {Pichon}, \& {Rhee}}]{Yi23}
{Yi}, S.~K., {Jang}, J.~K., {Devriendt}, J., {et~al.} 2023, arXiv e-prints,
  arXiv:2308.03566

\end{thebibliography}


\end{document}